\documentclass[twocolumn,showpacs,preprintnumbers,pra,amsmath,amssymb,superscriptaddress]{revtex4}

\usepackage{bm}
\newcommand{\bfsfI}{\mbox{\sffamily\bfseries{I}}}

\begin{document}
\title{Multipole interaction between atoms and their photonic
environment}
\date{Submitted to Phys. Rev. A April 1, 2003 \\
      Accepted for publication June 2, 2003}
\author{Martijn Wubs} \email{c.m.wubs@utwente.nl}
\homepage{http://tnweb.tn.utwente.nl/cops/} \affiliation{Complex
Photonic Systems, Faculty of Science and Technology, University of
Twente, P.O. Box 217, NL-7500~AE~~Enschede, The Netherlands}
\affiliation{Van der Waals-Zeeman Institute, University of
Amsterdam, Valckenierstraat 65, NL-1018 XE  Amsterdam, The
Netherlands}
\author{L.G.~Suttorp}
\affiliation{Institute for Theoretical Physics, University of
Amsterdam, Valckenierstraat 65, NL-1018 XE  Amsterdam, The
Netherlands}
\author{A.~Lagendijk}
\affiliation{Complex Photonic Systems, Faculty of Science and
Technology, University of Twente, P.O. Box 217,
NL-7500~AE~~Enschede, The Netherlands}
\begin{abstract}
Macroscopic field quantization is presented for a nondispersive
photonic dielectric environment, both in the absence and presence
of guest atoms. Starting with a minimal-coupling Lagrangian, a
careful look at functional derivatives shows how to obtain
Maxwell's equations before and after choosing a suitable gauge. A
Hamiltonian is derived with a multipolar interaction between the
guest atoms and the electromagnetic field. Canonical variables and
fields are determined and in particular the field canonically
conjugate to the vector potential is identified by functional
differentiation as minus the full displacement field. An important
result is that inside the dielectric a dipole couples to a field
that is neither the (transverse) electric nor the macroscopic
displacement field. The dielectric function is different from the
bulk dielectric function at the position of the dipole, so that
local-field effects must be taken into account.

\end{abstract}
\maketitle

\section{Introduction}\label{chapmuemudintro}

Optical properties of  atoms such as spontaneous-emission rates
can be strongly influenced by their dielectric environment
\cite{Purcell46}. It is well known that near a mirror, emission
rates can be enhanced or diminished, depending on the distance to
the mirror and the orientation of the atomic dipole moment
\cite{Drexhage70,Milonni94}. Inside optical  cavities, lifetime
effects are even stronger \cite{Kleppner81}. In three-dimensional
photonic crystals with a photonic band gap, spontaneous emission
would even be fully inhibited at any position in the crystal for
atomic transition frequencies within the band gap
\cite{Yablonovitch87}. Not only single-atom properties, but also
properties of two or several of atoms such as dipole-dipole
interactions and superradiance will be influenced by the
dielectric environment. Again, extreme changes compared to free
space can occur for atoms positioned inside microcavities
\cite{Kurizki96,Andrew00} or
 photonic crystals \cite{Kurizki90,John91}.

The above medium-modified processes must be described by a quantum
optical theory of dielectrics. In this paper such a theory will be
given of guest atoms interacting with a photonic dielectric
environment that is characterized by a given spatially varying and
real dielectric function $\varepsilon({\bf r})$. The guest atoms
by definition are the atoms which are  not included in the
dielectric function. These guests will be described
microscopically, whereas the dielectric is described
macroscopically in terms of the dielectric function only.

 Dielectric mirrors and photonic crystals are usually described by a frequency-independent
spatially varying refractive-index. Optical components such as
glass plates, lenses and optical cavities are some more examples.
In many cases, the refractive index can be considered as piecewise
constant, but not always: in  the so-called graded-index optical
fibres the refractive index in the core varies parabolically with
the radius \cite{Mynbaev01}. In this paper, the relative
dielectric function $\varepsilon({\bf r})$ is left unspecified
(but assume it to be piecewise continuously differentiable) so
that the theory describes both piecewise constant and continuously
varying dielectric functions.

The quantization of the electromagnetic field in free space can be
found in many textbooks on quantum optics
\cite{Loudon83,Craig84,Cohen87,Mandel95}. In
\cite{Knoell87,Glauber91,Dalton96,Dalton97} the more general
problem is addressed how to quantize the electromagnetic field in
a dielectric described by a given real dielectric function
$\varepsilon({\bf r})$ that depends on position.  The term
``macroscopic quantization'' has been coined for this procedure
\cite{Dalton96}. The special case of infinite photonic crystals is
treated in  \cite{Kweon95}.

Guest atoms can be described theoretically essentially in two
ways. The simplest way is to treat them as known probes of the
electromagnetic field in the medium. In that case, the
electromagnetic fields are found from Maxwell's equations in the
absence of the guest atoms; the guests are introduced as atoms
with given properties, such as transition frequencies and dipole
moments. The atoms are assumed to couple to the fields that were
found in their absence. The second and more fundamental way to
introduce guest atoms into the theory is to start with Maxwell's
equations that also contain as sources the charges that make up
the guest atoms. In this  second approach, a Hamiltonian for the
combined system of charges and fields should be found   that leads
to Maxwell's equations, both inside and outside the atoms.

Even for an atom in free space, the difference between these two
approaches has led to debates how to interpret the field to which
a dipole couples, either to the transverse part of the electric
field or to the displacement field
\cite{Power78,Power83,Babiker83,Ackerhalt84,Power85}. The latter
coupling is the correct (more fundamental) interpretation
\cite{Cohen87}, but for most observables there are no numerical
punishments when interpreting the field wrongly. However, inside a
dielectric, there would be a considerable difference between an
atom coupling to the electric field or to the displacement field.
In this paper it will be shown by using the second, more
fundamental way of introducing guest atoms, that neither dipole
coupling is correct in a dielectric. Moreover, the need to
consider local-field effects will arise in a na\-tural way.

Of course, one does not tell the whole truth about a dielectric
when describing it by a real, nondispersive and spatially varying
dielectric function $\varepsilon({\bf r})$. In a macroscopic
description one forgets details of the microscopic constituents of
the dielectric. Also, material dispersion and absorption of light
(transitions to nonradiative states in the dielectrics) are
neglected. It is well known that the dielectric function is a
response function that should be a dispersive and complex function
of frequency, so as to satisfy the Kramers-Kronig relations.
Certain sum rules \cite{Barnett98} for modified spontaneous
emission rates when a\-veraged over all frequencies will therefore
not hold in the present formalism that violates these relations.
The question is whether dispersion and absorption are important in
a particular experiment that one has in mind. Often optical
experiments are only interesting in a frequency range where
absorption is indeed negligible, for example when measuring light
emitted by excited atoms inside a photonic crystal. In such cases,
it is common practice to neglect material dispersion and
absorption in the theoretical description as well
\cite{Glauber91,Vats02}.

Quantum optical descriptions exist  where dielectric functions do
satisfy the Kramers-Kronig relations, both for homogeneous
\cite{Huttner92,Wubs01} and inhomogeneous dielectrics
\cite{Dung98,Scheel98,Savasta02,Dung02}.   Usually, the guest
atoms are introduced into these theories in the simplest of the
two ways described above, as probe atoms with known properties in
a two- or three-level description. It would be interesting to
introduce guest atoms in theories of inhomogeneous Kramers-Kronig
dielectrics in the more fundamental way, starting with Maxwell's
equations with the charges of the guest atoms as sources, but this
will not be done here. Nor will we look at more microscopic
descriptions of the dielectric \cite{Knoester89,Ho93,Juzeliunas96}
where dispersion shows up naturally. It would be very challenging
to derive optical predictions from a microscopic description of
light and the matter that builds up a complex dielectric such as a
photonic crystal.

The goal of this paper is to derive a Hamiltonian with multipolar
interaction between the guest atoms and the electromagnetic field
inside the inhomogeneous and nondispersive dielectric.  The dipole
Hamiltonian can then be found as an approximation. First, field
quantization of a dielectric without guest atoms is described in
Sec.~\ref{inhomwithout}. Atoms are introduced into the dielectric
in Sec.~\ref{inhomwithformalism}. The starting point will be
Maxwell's equations and a minimal-coupling Lagrangian that
produces these equations. Special attention is paid to check
whether Maxwell's equations still hold after choosing a gauge.
This requires an interesting analysis of functional
differentiation after choosing a gauge, presented in
Sec.~\ref{appfuncdef}. The minimal-coupling Lagrangian is then
transformed  in Sec.~\ref{secquantmultiham} to its multipolar form
with use of the Power-Zienau-Woolley transformation that is
well-known for free space \cite{Cohen87}. Our careful analysis of
functional differentiation allows us to use a transformation that
is simpler and more like the free-space case than presented in
related work \cite{Dalton96,Dalton97}.  After the transformation,
canonical variables and fields are determined. In particular, the
important question which field in the dielectric is canonically
conjugate to the vector potential can be answered more easily than
in \cite{Dalton96,Dalton97} and the answer will be different than
presented in \cite{Dalton96}. The multipolar Hamiltonian and its
dipole approximation are given in second-quantization notation.
Results are compared to the free-space case. The free-space
dipole-coupling controversy and confusion is reviewed
 in Sec.~\ref{dipolecontroversy} and the results of this paper are discussed in that perspective.
The quantum optics of dielectrics which also have inhomogeneous
magnetic properties are briefly discussed in
Sec.~\ref{magneticeffects}, before concluding in
Sec.~\ref{summarymuemudchap}.

\section{Inhomogeneous dielectric without guest
atoms}\label{inhomwithout}

\subsection{Classical Lagrangian and
Hamiltonian}\label{chapmuemudclasham} In this section
 the quantization of the electromagnetic field
in inhomogeneous dielectrics \cite{Knoell87,Glauber91} is
reviewed. The emphasis will be on concepts and results that will
be employed in the following sections, when guest atoms are
introduced in the dielectric.

 In SI-units, the source-free
Maxwell equations in matter  are \begin{subequations}
\begin{eqnarray}\label{Maxwell}
{\bm \nabla} \cdot {\bf B} & = & 0,  \qquad
{\bm \nabla} \times {\bf E} +  {\bf \dot{B}}  =  0, \\
{\bm \nabla} \cdot {\bf D} & = & 0,  \qquad
 {\bm \nabla} \times {\bf H} - {\bf \dot{D}}  =  0.
\label{Maxwell3}
\end{eqnarray}
\end{subequations}
Here and in the following the dot denotes a partial time
derivative. The fields ${\bf E}$, ${\bf B}$, ${\bf D}$ and ${\bf
H}$ are the electric field and the magnetic induction, the
displacement field and the magnetic field vector, respectively.
For nonmagnetic inhomogeneous dielectrics, the constitutive
relations are simply ${\bf B} = \mu_{0} {\bf H}$ and ${\bf D} =
\varepsilon_{0} \varepsilon({\bf r}){\bf E}$. (The generalization
of the present theory to inhomogeneous magnetic materials will be
discussed in section~\ref{magneticeffects}.) As for free space,
the electric and magnetic fields can be expressed in terms of a
vector potential ${\bf A}$ and a scalar potential $\Phi$:
\begin{equation}
{\bf E}  = -{\bm \nabla} \Phi -  {\bf \dot{A}},
  \qquad {\bf B} = {\bm \nabla} \times {\bf A}. \label{EBintermsofAPhi}
\end{equation}
There is gauge freedom in choosing pairs $({\bf A},\Phi)$ that
lead to the same electric and magnetic fields.  Now choose the
generalized Coulomb gauge which is defined by the requirement that
the vector potential satisfies
\begin{equation}\label{Agentrans}
{\bm \nabla} \cdot [\;\varepsilon({\bf r}){\bf A}({\bf r})\;] = 0.
\end{equation}
The vector potential or any field satisfying this condition, is
called ``generalized transverse'', because it satisfies a
generalized version of the Coulomb gauge condition \mbox{${\bm
\nabla} \cdot {\bf A}=0$} in free space. In the generalized
Coulomb gauge, the vector potential must satisfy the wave equation
\begin{equation}\label{waveA}
{\bm \nabla}\times{\bm \nabla}\times {\bf A}+
\frac{\varepsilon({\bf r})}{c^{2}}{\bf \ddot{A}}=0,
\end{equation}
in order to be consistent with the Maxwell equation
(\ref{Maxwell3}). The scalar potential can be chosen identically
zero ($\Phi\equiv 0$) in the generalized Coulomb gauge.

Since the goal is to find a quantum optical Hamiltonian in the
end, one should start with a Lagrangian for\-ma\-lism for the
classical Maxwell fields. From the Lagrangian the canonical fields
and their conjugates can be identified that will become pairs of
non-commuting field operators in a later stage.  The principle of
least action states that  fields (and particle variables, when
present) minimize the action \cite{Cohen87}; the action is defined
as the time-integrated Lagrangian between some initial and final
times. The requirement that small variations in the fields do not
change the action leads to the Euler-Lagrange equations for the
canonical fields, in our case for the vector potential
\cite{Cohen87,Craig84}:
\begin{equation}\label{EulerLagrange}
\frac{\delta L}{\delta {\bf A}}-
\frac{\mbox{d}}{\mbox{d}t}\frac{\delta L}{\delta \dot{\bf A}} = 0.
\end{equation}
Here, functional derivatives are denoted with ``$\delta$'' and
more will be said about them later.

 A Lagrangian $L_{0}$ for the electromagnetic field in an inhomogeneous medium is
\begin{equation}\label{lagrangiaglauber}
L_{0}= \int\mbox{d}{\bf r}\;\mathcal{L}_{0}\;\equiv
\frac{1}{2}\int\mbox{d}{\bf
r}\;\left[\varepsilon_{0}\varepsilon({\bf r})\dot{\bf A}^{2}-
\mu_{0}^{-1}({\bm \nabla}\times{\bf A})^{2}\right]. \end{equation}
The Lagrangian is the spatial integral of the Lagrangian density
$\mathcal{L}_{0}$ over a large volume $V$ that will eventually be
sent to infinity. The subscript ``$0$'' is used to denote the
absence of guest atoms in the dielectric. The vector potential is
a canonical field variable and its canonically conjugate field can
be found as a functional derivative of the Lagrangian density
\begin{equation}\label{canconj}
{\bm \Pi} \equiv \frac{\delta L_{0}}{\delta\dot{\bf A}}=
\varepsilon_{0}\varepsilon({\bf r}){\bf \dot{A}}= - {\bf D}.
\end{equation}
(The functional derivative is used somewhat naively here, but the
answer is correct, as a more detailed analysis in section
\ref{appfuncdef} will show.) In other words, the field canonically
conjugate to the vector potential equals minus the displacement
field, which is a transverse field. Proceeding as for free space
\cite{Cohen87}, one finds that the Euler-Lagrange equation of
motion for the vector potential leads to the wave equation
(\ref{waveA}) for the vector potential in the medium. The
Hamiltonian is
\begin{equation}\label{Hamiltonianepser}
H_{0}  =  \int \mbox{d}{\bf r}\; ({\bm \Pi}\cdot{\bf
\dot{A}}-\mathcal{L}_{0})  =  \frac{1}{2}\int \mbox{d}{\bf r}\;
\left[\frac{{\bm \Pi}^{2}}{\varepsilon_{0}\varepsilon({\bf
r})}+\frac{\left({\bm \nabla} \times {\bf
A}\right)^{2}}{\mu_{0}}\right].
\end{equation}
This is the Hamiltonian for the classical electromagnetic field in
an inhomogeneous dielectric, without  guest atoms.

\subsection{Complete sets and quantum
Hamiltonian}\label{completesetsandquantum}
 For a quantum optical
description of the dielectric, the electromagnetic fields can best
be expanded in terms of harmonic solutions of the wave equation
(\ref{waveA}). With each of these ``true modes'' one can associate
independent canonical variables, for which commutation rules can
be given. The set of true modes is not unique. This freedom will
be used below to choose a particularly convenient set.  For
example, in vacuum the true modes are transverse plane waves. For
the plane waves, one can choose linear combinations of cosine and
sine solutions $\cos({\bf k}\cdot{\bf r})$ and $\sin({\bf
k}\cdot{\bf r})$. The complex exponential $\exp(i {\bf k}\cdot{\bf
r})$ is only one such linear combination.

 It will now be argued why the field can be expanded in terms of
a set of real mode functions, and the orthonormality relations of
the modes will be derived. The choice of real modes will simplify
the quantization procedure. The reason not to start  with complex
mode functions is that the latter has associated generalized
coordinates and momenta that are not Hermitian. This makes
intermediate results more complicated
\cite{Glauber91,Vogel94,Dalton96}. Real mode functions were also
used in \cite{Knoell87}, without the motivation given here.

Why does a complete set of real mode functions exist? As before,
the electromagnetic fields are assumed to live in the large volume
$V$. Let $Q$ be an abstract operator in Hilbert space which has a
local representation $\langle {\bf r}|Q|{\bf r'}\rangle =
\delta({\bf r}-{\bf r'}) Q({\bf r})$ in configuration space:
\begin{equation}\label{Qdef}
  Q({\bf r}) \equiv \frac{1}{\sqrt{\varepsilon({\bf
r})}}{\bm \nabla}\times{\bm \nabla}\times
\frac{1}{\sqrt{\varepsilon({\bf r})}}.
\end{equation}
The operator $Q$ is Hermitian under the normal inner product.
Eigenvalues of $Q$ are $(\omega_{\lambda}/c)^{2}$ and the
$\omega_{\lambda}$ will be called eigenfrequencies. All
eigenfunctions ${\bf g}_{\lambda}$ of $Q$ have the property ${\bm
\nabla}\cdot[\sqrt{\varepsilon({\bf r})}{\bf g}({\bf r})]=0$. The
label ${\lambda}$ is understood to count both continua and
discrete sets of solutions.
 The subspace of functions in Hilbert space
with the same transversality property is spanned by the
eigenfunctions of $Q$. Now let $C$ be the operator which is also
local in configuration space, where its  action  is to take the
complex conjugate. The dielectric function in this context must
also be viewed as an abstract operator $\varepsilon$ with local
representation in configuration space: $\langle {\bf
r}|\varepsilon|{\bf r'}\rangle= \delta({\bf r}-{\bf
r'})\varepsilon({\bf r})$. The representations of $Q$ and $C$
commute in configuration space, because $\varepsilon({\bf r})$ is
real for all positions ${\bf r}$ in $V$. Then $Q$ and $C$ commute
in any representation. From the fact that $Q$ and $C$ commute it
follows that an orthonormal basis of real eigenfunctions $\{{\bf
g}_{\lambda}\}$ of $Q$ can be chosen to span the subspace (with
complex coefficients).

The above analysis shows that the vector potential ${\bf A}$ can
be expanded in terms of a complete set of {\em real} vector mode
functions $\{{\bf h}_{\lambda}({\bf r})\}\equiv
\{\sqrt{\varepsilon({\bf r})}{\bf g}_{\lambda}({\bf r})\}$, which
are the harmonic solutions of the wave equation (\ref{waveA}):
\begin{equation}\label{fkwave}
{\bm \nabla}\times{\bm \nabla}\times {\bf h}_{\lambda}({\bf r}) -
\frac{\varepsilon({\bf r})\omega_{\lambda}^{2}}{c^{2}}{\bf
h}_{\lambda}({\bf r})=0. \end{equation} These mode functions
satisfy the same generalized transversality condition
(\ref{Agentrans}) as the vector potential.
 As is clear from Eq.~(\ref{fkwave}), unlike the
${\bf g}_{\lambda}$ the functions ${\bf h}_{\lambda}$ do not
satisfy a Hermitian eigenvalue equation. (To each type of modes
corresponds a different density of states \cite{Lagendijk96}.)
From the orthonormality of the ${\bf g}_{\lambda}$ it follows that
the functions ${\bf h}_{\lambda}({\bf r})$ satisfy the generalized
orthonormality condition \cite{Knoell87,Glauber91}
\begin{equation}\label{finnerproduct}
\int\mbox{d}{\bf r}\;\varepsilon({\bf r}){\bf
h}^{*}_{\lambda}({\bf r})\cdot {\bf h}_{\lambda'}({\bf r}) =
\delta_{\lambda \lambda'}.
\end{equation}
The complex-conjugation symbol $*$ was written for future
reference, since of course ${\bf h}_{\lambda}$ is real.  The
equation (\ref{finnerproduct}) will be called a generalized inner
product of the modes ${\bf h}_{\lambda}$ and ${\bf h}_{\lambda'}$.
The spatial integral in (\ref{finnerproduct}) runs over the volume
$V$,  so that the mode functions scale as $(V)^{-1/2}$.

In a scattering situation, where $\varepsilon({\bf r})$ is a
space-filling dielectric function plus a local modification within
a scattering volume $V_{\rm s}$, the contribution of the
scattering volume to the integral (\ref{finnerproduct}) scales as
$V_{\rm s}/V$. This fraction becomes of measure zero when the
quantization volume $V$ is sent to infinity. As an example, mode
functions of an infinite photonic crystal with a single point
defect (extra or missing dielectric material) have the same
orthonormality relations as the  mode functions in the absence of
the defect.

 The functions ${\bf h}_{\lambda}$ are complete in
the sense that they form a basis for generalized transverse
functions (such as the vector potential) that satisfy the wave
equation (\ref{waveA}). In other words, a generalized transverse
delta function ${\bm \delta}_{\varepsilon}^{\rm T}$ (a
distribution) can be defined in terms of the functions ${\bf
h}_{\lambda}$:
\begin{equation}\label{deltagegtrans}
{\bm \delta}_{\varepsilon}^{\rm T}({\bf r},{\bf r'}) \equiv
\sum_{\lambda} {\bf h}_{\lambda}({\bf r}){\bf h}_{\lambda}({\bf
r'})\varepsilon({\bf r'}).
\end{equation}
For $\varepsilon({\bf r})\equiv 1$, this expression reduces to the
(real) free-space transverse delta function (see \cite{Craig84},
p.~53)
\begin{equation}\label{deltatransfree}
{\bm \delta}^{\rm T}({\bf r})  =  \frac{2}{3}\delta({\bf r})\bfsfI
- \frac{1}{4 \pi r^{3}}(\bfsfI-3 \hat{\bm r}\otimes \hat{\bm r}),
\end{equation}
where $\bfsfI$ is the unit tensor in three dimensions and
$\hat{\bm r}$ is the unit vector ${\bf r}/|{\bf r}|$.
 Evidently, the generalized transverse delta
function (\ref{deltagegtrans}) is real because the mode functions
are real. From its  definition (\ref{deltagegtrans}) and the
normalization condition (\ref{finnerproduct}) of the modes, it
follows that ${\bm \delta}_{\varepsilon}^{\rm T}$ is idempotent:
\begin{equation}\label{deltaidempotent}
\int\mbox{d}{\bf r}_{1}\;{\bm \delta}_{\varepsilon}^{\rm T}({\bf
r},{\bf r}_{1})\cdot {\bm \delta}_{\varepsilon}^{\rm T}({\bf
r}_{1},{\bf r'})= {\bm \delta}_{\varepsilon}^{\rm T}({\bf r},{\bf
r'}),
\end{equation} In other words, ${\bm \delta}_{\varepsilon}^{\rm T}$ is a projector
into the subspace of generalized transverse functions. The
generalized transverse delta function is not symmetric in its
arguments, because it is transverse in its second and generalized
transverse in its first variable:
\begin{equation}\label{transdeltaisgentrans}
{\bm \nabla}_{\bf r}\cdot\left[\varepsilon({\bf r}){\bm
\delta}_{\varepsilon}^{\rm T}({\bf r},{\bf r'})\right] =0={\bm
\nabla}_{\bf r'}\cdot\left[{\bm \delta}_{\varepsilon}^{\rm T}({\bf
r},{\bf r'})\right].
\end{equation}

The vector potential and its canonically conjugate field have
normal-mode expansions
\begin{subequations}
\begin{eqnarray}\label{Apinormalmode}
{\bf A}({\bf r},t) & = &
1/(\sqrt{\varepsilon_{0}})\sum_{\lambda}q_{\lambda}(t)\;{\bf
h}_{\lambda}({\bf r}), \label{Anormalmode}\\
{\bm \Pi}({\bf r},t) & = & \sqrt{\varepsilon_{0}}
\sum_{\lambda}p_{\lambda}(t)\;\varepsilon({\bf r}){\bf
h}_{\lambda}({\bf r}),\label{pinormalmode}
\end{eqnarray}
\end{subequations}
with  generalized coordinates $q_{\lambda}(t)$ and momenta
$p_{\lambda}(t)$. At this point the choice of real mode functions
pays off,  because the associated generalized coordinates and
momenta are only real when the modes are real; only real
coordinates and momenta will become Hermitian operators in a
quantum description. If the above two expansions are substituted
in the Hamiltonian (\ref{Hamiltonianepser}) and the normalization
condition (\ref{finnerproduct}) is used,  then it follows that $H
= \frac{1}{2}\sum_{\lambda}(p_{\lambda}^{2} +
\omega_{\lambda}^{2}q_{\lambda}^{2})$. The Hamiltonian turns out
to be a simple sum over the true modes of the inhomogeneous
dielectric, where the energy of each mode corresponds to a
one-dimensional harmonic oscillator with position $q_{\lambda}$,
momentum $p_{\lambda}$ and frequency $\omega_{\lambda}$.

Now comes the quantization step. The independent canonical pairs
satisfy the standard equal-time commutation relations
$\left[q_{\lambda}(t), p_{\lambda'}(t)\right] = i \hbar
\delta_{\lambda\lambda'}$. With the normal mode expansions
(\ref{Anormalmode})  and (\ref{pinormalmode}), the commutation
relation for the vector potential and its canonically conjugate
field can be found immediately:
\begin{equation}\label{cancomapi}
\left[ {\bf A}({\bf r},t), {\bm \Pi}({\bf r'},t)\right] = i \hbar
\sum_{\lambda} {\bf h}_{\lambda}({\bf r}){\bf h}_{\lambda}({\bf
r'})\varepsilon({\bf r'})= i \hbar {\bm \delta}_{\varepsilon}^{\rm
T}({\bf r},{\bf r'}).
\end{equation}
The commutator is a dyadic quantity. It turns out to be
proportional to the generalized transverse delta function.

Annihilation operators are introduced as $\alpha_{\lambda}  =
\sqrt{\omega_{\lambda}/(2 \hbar)}q_{\lambda} + i \sqrt{1/(2 \hbar
\omega_{\lambda})}p_{\lambda}$ and creation operators as their
Hermitian conjugates. They have standard commutation relations
$[\alpha_{\lambda}(t),\alpha_{\lambda'}^{\dag}(t)]=\delta_{\lambda\lambda'}$
and all other inequivalent commutators are zero. The Hamiltonian
becomes the sum over contributions
$\hbar\omega_{\lambda}(\alpha_{\lambda}^{\dag}\alpha_{\lambda}+1/2)$
of individual modes. Thus the concept of a photon as the
elementary excitation $\alpha^{\dag}_{\lambda}|0\rangle$ of a mode
is as useful for inhomogeneous dielectrics as it is for free
space. Number states, coherent and squeezed states etcetera can be
defined analogously. The only difference for inhomogeneous
dielectrics is that their true modes are not plane waves.

The vector potential operator and its canonically conjugate field
operator can be expressed in terms of creation and annihilation
operators as
\begin{subequations}
\begin{eqnarray}\label{Apinormalmode2}
{\bf A}({\bf r},t) & = & \sum_{\lambda} \sqrt{\frac{\hbar}{2
\varepsilon_{0} \omega_{\lambda}}}\left[ \alpha_{\lambda}(t)\;{\bf
h}_{\lambda}({\bf r}) + {\rm H.c.}\right], \label{Anormalmode2}\\
{\bm \Pi}({\bf r},t) & = & - i \varepsilon_{0}\varepsilon({\bf r})
\sum_{\lambda}\sqrt{\frac{\hbar\omega_{\lambda}}{2\varepsilon_{0}}}
\left[ \alpha_{\lambda}(t)\;{\bf h}_{\lambda}({\bf r}) - {\rm
H.c.} \right].\label{pinormalmode2}
\end{eqnarray}
\end{subequations}
Here, ``H.c.'' denotes the Hermitian conjugate. The forms of the
electric and magnetic fields as quantum mechanical operators as
well as their  commutation relations immediately follow from
(\ref{EBintermsofAPhi}) and the above equation
(\ref{Anormalmode2}). The time-dependence of the operators is
simply harmonic, for example
$\alpha_{\lambda}(t)=\alpha(0)\exp(-i\omega_{\lambda}t)$.

In practice, it can be convenient to use a set of complex true
mode functions $\{ {\bf f}_{\mu}\}$ instead of the real mode
functions $\{ {\bf h}_{\lambda}\}$. Since the complex mode
functions should also satisfy the wave equation (\ref{fkwave}) and
the generalized orthonormality condition (\ref{finnerproduct}),
the two sets of mode functions are related through a unitary
transformation ${\bf f}_{\mu} = \sum_{\lambda} U_{\mu
\lambda}\;{\bf h}_{\lambda}$ that only relates mode functions with
identical eigenfrequencies; ${\bf U}$ is a unitary matrix. Note
that because of this unitary relation, the generalized transverse
delta function (\ref{deltagegtrans}) can alternatively be
expressed in terms of the complex mode functions ${\bf
f}_{\lambda}$. Its effect is the substitution of one ${\bf
h}_{\lambda}$ by ${\bf f}_{\lambda}$ and the other ${\bf
h}_{\lambda}$ by ${\bf f}_{\lambda}^{*}$: ${\bm
\delta}_{\varepsilon}^{\rm T}({\bf r},{\bf r'})$ is also equal to
$\sum_{\lambda}{\bf f}_{\lambda}({\bf r}){\bf
f}^{*}_{\lambda}({\bf r'})\varepsilon({\bf r'})$.
 However, after this substitution  it is no longer obvious that
  ${\bm \delta}_{\varepsilon}^{\rm T}({\bf r},{\bf r'})$ is real-valued.

The  field operators can also be expanded in terms of the complex
mode functions ${\bf f}_{\lambda}$ as
\begin{subequations}
\begin{eqnarray}\label{Apinormalmode3}
{\bf A}({\bf r},t) & = & \sum_{\lambda} \sqrt{\frac{\hbar}{2
\varepsilon_{0}
\omega_{\lambda}}}\left[\;a_{\lambda}^{(0)}(t)\;{\bf
f}_{\lambda}({\bf r}) + {\rm H.c.}\right], \label{Anormalmode3}\\
{\bm \Pi}({\bf r},t) & = & - i \varepsilon_{0}\varepsilon({\bf r})
\sum_{\lambda}\sqrt{\frac{\hbar\omega_{\lambda}}{2\varepsilon_{0}}}
\left[\; a_{\lambda}^{(0)}(t)\;{\bf f}_{\lambda}({\bf r}) - {\rm
H.c.} \right],\label{pinormalmode3}
\end{eqnarray}
\end{subequations}
where the new  annihilation operator $a_{\mu}^{(0)}$  associated
with the complex mode ${\bf f}_{\mu}$ is defined in terms of the
``old'' operators as $a_{\mu}^{(0)} \equiv \sum_{\lambda}
U_{\mu\lambda}^{-1}\;\alpha_{\lambda}$, and  $a_{\mu}^{(0)\dag}$
is its Hermitian conjugate. The commutation relations of
$a_{\mu}^{(0)}$ and $a_{\mu}^{(0)\dag}$ are again the standard
relations, because ${\bf U}$ is a unitary transformation. To
distinguish $a_{\mu}^{(0)}$ from  operators to be defined later,
the superscript $(0)$ has been added, signifying that no guest
atoms are present. The time dependence of the operators is again
harmonic.

This completes the quantization of the electromagnetic field in an
inhomogeneous dielectric without guest atoms. The reason  to start
the quantization procedure with real mode functions was that the
associated generalized coordinates and momenta are real quantities
that  become Hermitian operators in quantum mechanics. It is
possible to start  with complex mode functions instead and to
proceed with the non-Hermitian operators
\cite{Glauber91,Vogel94,Dalton96}, but it makes intermediate
results unnecessarily more complicated.  The unitary relations
between complex and real mode functions and between their
respective annihilation operators are purely formal, unless both
sets of mode functions are given explicitly. The above
quantization procedure only relies on the mere existence (rather
than on an explicit construction) of these unitary mappings.

\section{Inhomogeneous dielectric with guest
atoms}\label{inhomwithformalism}

In the previous section it was described how to quantize the
electromagnetic field in an inhomogeneous dielectric. Now inside
the inhomogeneous dielectric guest atoms are introduced.  Their
optical response is not included in the dielectric function
$\varepsilon({\bf r})$ of the medium. The goal in the following
sections is to find the quantum optical description of the
combined system, with a multipole interaction between the
electromagnetic field and the guest atoms. There are at least two
reasons why the multipolar Hamiltonian is to be preferred. In the
first place, it is more convenient when only approximate
calculations can be done which in the minimal-coupling formalism
would give gauge-dependent results \cite{Loudon83}; secondly,
atoms are much smaller than optical wavelengths and in the
multipole-formalism this can be exploited well. Actually, atoms
are so much smaller than optical wavelengths that often ``atoms''
are identified with ``dipoles''.

The starting point is the minimal-coupling Lagrangian  that
produces the Maxwell equations and the equations of motion for the
charges that make up the guest atoms. The minimal-coupling
Lagrangian can be used to find a minimal-coupling Hamiltonian and
this procedure can be found in \cite{Knoell87,Kweon95}. A clear
exposition is also given in \cite{Hooijer01}. Here the Lagrangian
will first be transformed to the multipolar form before
constructing a Hamiltonian. The latter procedure was followed also
in \cite{Dalton96,Dalton97}. The present work is different in some
essential aspects that will be stressed where appropriate.

\subsection{Choice of suitable Lagrangian}
Guest atoms inside an inhomogeneous dielectric can be described by
a charge density $\sigma_{\rm g}$ and a current density ${\bf
J}_{\rm g}$ which show up as sources in Maxwell's equations
\cite{Loudon83,Dalton96}:
\begin{subequations}
\begin{eqnarray}\label{maxwelldielectricandguests}
{\bm \nabla}\cdot {\bf B} & = & 0, \label{Maxwellfirst} \\ {\bm
\nabla} \times {\bf E} + {\bf \dot{B}} & = & 0,
\label{Maxwellsecond}
\\
\varepsilon_{0}{\bm \nabla}\cdot \left[\varepsilon({\bf r}){\bf
E}({\bf r})\right] & = & \sigma_{\rm g},  \label{Maxwellthird} \\
\mu_{0}^{-1}{\bm \nabla} \times {\bf B} -
\varepsilon_{0}\varepsilon({\bf r}){\bf \dot{E}}  & = &  {\bf
J}_{\rm g}. \label{Maxwellfourth}
\end{eqnarray}
\end{subequations}
Here,  $\sigma_{\rm g}$ is the charge density and ${\bf J}_{\rm
g}$ the current density produced by the guest atoms alone, as
stressed by the subscript ``${\rm g}$''; the dielectric is
completely described by the dielectric function
$\varepsilon_{0}\varepsilon({\bf r})$ and the magnetic
permeability $\mu_{0}$. Whatever Lagrangians and Hamiltonians are
introduced for the inhomogeneous dielectric plus guest atoms, they
must lead to these four Maxwell  equations. Moreover, the
electrons with charges $-e$ and masses $m_{\rm e}$ should respond
to electric and magnetic fields as given in the equation of motion
\begin{equation}\label{chargefeelsfields}
m_{\rm e}{\bf \ddot{r}}_{mj} = -e\left[ {\bf E}({\bf r}_{mj}) +
{\bf \dot{r}}_{mj}\times{\bf B}({\bf r}_{mj})\right].
\end{equation}
We assume that there are no free charges. All electrons (labelled
$j$) are bound to atomic nuclei (label $m$) to form neutral guest
atoms. Then $\sigma_{\rm g}$ and ${\bf J}_{\rm g}$ are given by
\cite{Jackson75}
\begin{subequations}
\begin{eqnarray}\label{chargecurrent}
\sigma_{\rm g}({\bf r},t) & = & e\sum_{m}\left[ Z_{m}\delta({\bf
r}-{\bf R}_{m})-\sum_{j}\delta({\bf r}-
{\bf r}_{mj})\right] \label{chargecurrentcharge}\\
{\bf J}_{\rm g}({\bf r},t) & = & -e\sum_{m}\left[\sum_{j}
\dot{{\bf r}}_{mj} \delta({\bf r}- {\bf r}_{mj})\right].
\label{chargecurrentcurrent}
\end{eqnarray}
\end{subequations}
Here, $Z_{m}$ is the nuclear charge of atom $m$.  The guest atoms
are assumed to have fixed positions, their nuclei are their
centers of mass and are stationary at positions ${\bf R}_{m}$.
From these explicit forms of $\sigma_{\rm g}$ and ${\bf J}_{\rm
g}$  follows the equation of continuity or current conservation,
${\bm \nabla}\cdot {\bf J}_{\rm g}+\dot{\sigma}_{\rm g}=0$, which
can also be found from the Maxwell equations (\ref{Maxwellthird})
and (\ref{Maxwellfourth}).

Again, the electric and magnetic fields can be defined through
Eq.~(\ref{EBintermsofAPhi}) in terms of a vector potential ${\bf
A}$ and a scalar potential $\Phi$. Then the  two homogeneous
Maxwell equations (\ref{Maxwellfirst}) and (\ref{Maxwellsecond})
are automatically satisfied. The other two Maxwell equations
should follow from the Euler-Lagrange equations
[Eq.~(\ref{EulerLagrange})] for the scalar and the vector
potential, respectively. The minimal-coupling Lagrangian is
\begin{equation}\label{Lagrangianchoice}
L_{\rm min} = \sum_{m,j}\frac{1}{2}m_{\rm e}{\bf \dot{r}}_{mj}^{2}
+ \int\mbox{d}{\bf r}\;\mathcal{L}_{\rm min}.
\end{equation}
Here the Lagrangian density $\mathcal{L}_{\rm min}$  describes the
electromagnetic field energy and its minimal-coupling interaction
with the guest atoms:
\begin{equation}\label{Lagdensity}
\mathcal{L}_{\rm min}= \frac{1}{2}\varepsilon_{0}\varepsilon({\bf
r})\left[ {\bf \dot{A}}+{\bm \nabla}\Phi\right]^{2}-
\frac{1}{2\mu_{0}}({\bm \nabla}\times {\bf A})^{2} + {\bf J}_{\rm
g}\cdot{\bf A}-\sigma_{\rm g}\Phi.
\end{equation}
Indeed, Maxwell's third and fourth equations can be found from the
Euler-Lagrange equations  for the scalar and the vector
potentials, respectively. Moreover, the Euler-Lagrange equations
for the canonical variables ${\bf r}_{mj}$ give the equations of
motion (\ref{chargefeelsfields}) for the charged particles. Note
that the Lagrangian leads to these equations of motion, before
choosing a gauge to fix ${\bf A}$ and $\Phi$ with: the equations
of motion are gauge-independent results that should not depend on
the choice of gauge.

\subsection{Fixing the gauge}\label{fixinggauge}
 The electric and magnetic fields are
defined in terms of a scalar and a vector potential. But there is
gauge freedom, which  means that  the scalar and vector potentials
are not uniquely defined by the requirement that measurable
electric and magnetic fields satisfy Maxwell's equations. We need
to choose a gauge in order to find in the end a quantum mechanical
description of light interacting with the guest atoms.  As in the
situation without guest atoms in section \ref{inhomwithout}, the
generalized Coulomb gauge is chosen so that the vector potential
satisfies  Eq.~(\ref{Agentrans}). In this section it will be
checked whether the equations of motion for the scalar and vector
potentials still lead to the third and fourth Maxwell equations
after choosing the gauge. This must be the case, because the
choice of gauge should not change the physical predictions of the
theory. Still, the check was not performed in
\cite{Dalton96,Dalton97} and, as we shall see, it will be very
useful to do so here.

The gauge affects the interaction term
\begin{equation}\label{LAPhi}
L_{{\bf A}\Phi}\equiv \varepsilon_{0}\int\mbox{d}{\bf
r}\;\varepsilon({\bf r}){\bf \dot{A}}\cdot{\bm \nabla}\Phi,
\end{equation}
of the Lagrangian (\ref{Lagrangianchoice}). The term becomes
identically zero, because in the generalized Coulomb gauge it has
become an inner product of a transverse and a longitudinal
function. The remaining terms in the Lagrangian involving the
scalar potential lead to an Euler-Lagrange equation that is the
generalized Poisson equation for the scalar potential in the
Coulomb gauge:
\begin{equation}\label{generalpoisson}
\varepsilon_{0}{\bm \nabla}\cdot\left[\varepsilon({\bf r}){\bm
\nabla}\Phi({\bf r})\right] = -\sigma_{\rm g}.
\end{equation}
Clearly, the scalar potential can not be chosen identically zero
as in the situation without guest atoms. The gauge-fixing
condition (\ref{Agentrans}) for the vector potential, the equation
(\ref{generalpoisson}) for the scalar potential, together with the
definition of the electric field (\ref{EBintermsofAPhi}) in terms
of the two potentials, still lead to the third Maxwell equation
(\ref{Maxwellthird}).

As in free space, the scalar potential is a function of the
positions of the charges that make up the guest atoms
\cite{Cohen87}. In other words, one can first solve the coupled
equations of motion for the vector potential and the charges, and
from the charge distribution $\sigma_{\rm g}(t)$ thus found, the
scalar potential $\Phi(t)$ can be found as the solution of
Eq.~(\ref{generalpoisson}). Therefore,  the scalar potential is
not an independent canonical field. The Lagrangian
(\ref{Lagrangianchoice}) can then be simplified as
\begin{equation}\label{Lrewritten}
{L'}_{\rm min} = \sum_{m,j}\frac{1}{2}m_{\rm e}{\bf
\dot{r}}_{mj}^{2} -V_{C} +\int\mbox{d}{\bf r}\;\mathcal{L'}_{\rm
min},
\end{equation} where Eqs.~(\ref{LAPhi}) and (\ref{generalpoisson})
were used. The Coulomb interaction $V_{C} =
(\varepsilon_{0}/2)\int\mbox{d}{\bf r}\;\varepsilon({\bf r})({\bm
\nabla} \Phi)^{2}$ is a function of the guest atoms alone; the
Lagrangian density in (\ref{Lrewritten})  becomes
\begin{equation}\label{Lagdensitynew}
\mathcal{L'}_{\rm min} =
\frac{1}{2}\varepsilon_{0}\varepsilon({\bf r}){\bf
\dot{A}}^{2}-\frac{1}{2\mu_{0}}({\bm \nabla}\times {\bf A})^{2} +
{\bf J}_{\rm g}\cdot{\bf A}.
\end{equation}
Which equation do we find for the vector potential after choosing
the generalized Coulomb gauge? Let us begin at the other end: in
order to be consistent with the fourth Maxwell equation
(\ref{Maxwellfourth}), the vector potential should satisfy
\begin{equation}\label{waveforA}
\mu_{0}^{-1}{\bm \nabla}\times{\bm \nabla}\times {\bf A}+
\varepsilon_{0}\varepsilon({\bf r}) {\bf \ddot{A}} = {\bf J}_{\rm
g}-\varepsilon_{0}\varepsilon({\bf r}){\bm \nabla} \dot{\Phi}.
\end{equation}
It is not obvious how  the source term
$-\varepsilon_{0}\varepsilon({\bf r}){\bm \nabla} \dot{\Phi}$ can
appear at the right-hand side of this equation by functional
differentiation of the Lagrangian with respect to the vector
potential. Before choosing the gauge, this source term originated
from the interaction term $L_{{\bf A}\Phi}$ [Eq.~(\ref{LAPhi})],
which is zero after choosing the gauge.  After choosing the
generalized Coulomb gauge, a more careful analysis is needed in
order to find the fourth Maxwell equation.

\section{Functional differentiation after choosing the gauge}
\label{appfuncdef}
 After choosing the gauge, the vector potential is generalized transverse.
 The Euler-Lagrange
equation for the vector potential is therefore an equation of
motion of a constrained system, where the constraint is the gauge
condition~(\ref{Agentrans}). One could try and solve this problem
using the method of Lagrange multipliers~\cite{Itzykson80}, but
this is not the route that will be pursued here. Instead, the
appropriate mathematical definition and computation of functional
derivatives after choosing a gauge  will be studied in the
following  subsections~\ref{funcdefdef} and \ref{simplerules},
respectively. The results will be applied to our physical problem
in subsection~\ref{funcdifminimal}.

\subsection{Two definitions of functional
derivatives}\label{funcdefdef} Let us generalize the problem
somewhat by considering a functional $F = \int\mbox{d}{\bf
r}\;\mathcal{F}$ with a functional density $\mathcal{F}$ that
depends on the three-dimensional  vector fields  ${\bf X}$ and
${\bf Y}$. Assume also that at some stage the generalized Coulomb
gauge will be chosen for the field ${\bf Y}$. This gauge is
defined by the requirement that ${\bm \nabla}_{\bf
r}\cdot[\varepsilon({\bf r}){\bf Y}({\bf r})]$ equals zero.

Before choosing the gauge, the functional derivative of the
functional $F$ with respect to the vector field ${\bf Y}$ is
defined as
\begin{equation}\label{funcdifdef}
\frac{\delta F}{\delta {\bf Y}({\bf r})} \equiv  \lim_{\gamma
\rightarrow 0} \frac{\int\mbox{d}{\bf r'}\;\{ \mathcal{F}[{\bf
Y}({\bf r'}) + \gamma\; \delta({\bf r}-{\bf
r'})\bfsfI\;]-\mathcal{F}[{\bf Y}({\bf r'})] \}}{\gamma}.
\end{equation}
(The ${\bf X}$-dependence of $\mathcal{F}$ was dropped for
brevity.) The functional derivative of  $F$ with respect to  ${\bf
Y}$ describes the relative changes of $F$ when small variations
proportional to $\delta({\bf r}-{\bf r'})\bfsfI$ are added to the
vector function ${\bf Y}$. Here, $\delta({\bf r}-{\bf r'})$ is the
Dirac delta function in three dimensions; as before, $\bfsfI$ is
the unit tensor. It turns out that the right-hand side of
Eq.~(\ref{funcdifdef}) can be computed as the partial derivative
of the functional density $\mathcal{F}$ with respect to ${\bf Y}$.
While doing this, the  $\mathcal{F}$ can simply be considered as a
function and ${\bf Y}$ as one of its variables. Before choosing
the generalized Coulomb gauge, Eq.~(\ref{funcdifdef}) correctly
defines the functional derivative of $F$ with respect to ${\bf
Y}$.

Now suppose for the moment that $F$ is defined as $F =
\int\mbox{d}{\bf r}\;{\bf X}\cdot{\bf Y}$. Suppose also that the
field ${\bf X}$ is the product of $\varepsilon({\bf r})$ with some
longitudinal vector field. Then $F$ becomes identically zero in
the generalized Coulomb gauge, because it is the inner product of
a transverse and a longitudinal vector field. Still, the
functional derivative Eq.~(\ref{funcdifdef}) of $F$ would give a
nonzero answer. This can only mean that Eq.~(\ref{funcdifdef})
does not define the functional derivative with respect to
generalized transverse functions correctly. The reason is that the
function space in which the field ${\bf Y}$ lives has become
smaller by choosing the gauge: it now lives in the subspace of
functions which are generalized transverse. This also means that
functional variations of ${\bf Y}$ should stay inside this
subspace. In the functional derivative (\ref{funcdifdef}),
variations in the whole function space are allowed and clearly
${\bm \nabla}_{\bf r'}\cdot[\varepsilon({\bf r'}) \delta({\bf
r}-{\bf r'})\bfsfI\;]$ is nonzero.

In general, with every set of  constraints  a new functional
derivative can be associated. Here, only the gauge-constraint will
be considered that functions be generalized transverse. Functional
differentiation with respect to generalized transverse functions
can be defined as (see \cite{Vogel94}, p.~20)
\begin{equation}\label{funcdifdefgentrans}
\frac{\delta F}{\delta {\bf Y}_{\varepsilon}^{\rm T}({\bf r})}
\equiv \lim_{\gamma \rightarrow 0} \frac{\int\mbox{d}{\bf r'}\;\{
\mathcal{F}[{\bf Y}({\bf r'}) + \gamma\; {\bm
\delta}_{\varepsilon}^{\rm T}({\bf r'},{\bf r})]-\mathcal{F}[{\bf
Y}({\bf r'})] \}}{\gamma},
\end{equation}
with the generalized transverse delta function ${\bm
\delta}_{\varepsilon}^{\rm T}$ as defined in
Eq.~(\ref{deltagegtrans}). In this new functional derivative, the
functional variations do stay inside the generalized transverse
subspace, since ${\bm \delta}_{\varepsilon}^{\rm T}$ is the
projector into the subspace and  ${\bm \nabla}_{\bf
r'}\cdot[\varepsilon({\bf r'}) {\bm \delta}_{\varepsilon}^{\rm
T}({\bf r'},{\bf r})]=0$. The derivative
(\ref{funcdifdefgentrans}) will be called the ``constrained
functional derivative'' in the following.

\subsection{Simple rules to compute constrained functional
derivatives}\label{simplerules} Now the goal is to find simple
rules to compute the constrained functional derivative
(\ref{funcdifdefgentrans}) with respect to generalized transverse
functions, just like  the normal functional derivative
(\ref{funcdifdef}) can simply be calculated as a partial
derivative. With that goal in mind, first some properties of
generalized transverse functions are derived.

With every transverse function ${\bf X}^{\rm T}({\bf r})$ a
generalized transverse function $[{\bf X}^{\rm T}({\bf r})
/\varepsilon({\bf r})]$ can be associated. From section
\ref{completesetsandquantum} we know that the latter function has
an expansion in terms of generalized transverse modes ${\bf
h}_{\lambda}({\bf r})$,  so that ${\bf X}^{\rm T}({\bf r})$ can be
expanded in terms of $\varepsilon({\bf r}){\bf h}_{\lambda}({\bf
r})$. This simple fact, in combination with
Eqs.~(\ref{finnerproduct}) and (\ref{deltagegtrans}), leads to the
following projection properties of ${\bm
\delta}_{\varepsilon}^{\rm T}$:
\begin{subequations}
\begin{eqnarray}\label{propsdeltaeps}
  \int\mbox{d}{\bf r'}\;{\bf X}^{\rm
T}({\bf r'})\cdot  {\bm \delta}_{\varepsilon}^{\rm T}({\bf
r'},{\bf r}) & = &   {\bf X}^{\rm T}({\bf r}),
\label{propsdeltaeps1}
\\
  \varepsilon({\bf
r})\int\mbox{d}{\bf r'}\;{\bm \delta}_{\varepsilon}^{\rm T}({\bf
r},{\bf r'})\cdot{\bf X}^{\rm T}({\bf r'})/\varepsilon({\bf r'})
 & = & {\bf X}^{\rm T}({\bf r}), \label{propsdeltaeps1b}
\\
 \int\mbox{d}{\bf r'}\; {\bm \delta}_{\varepsilon}^{\rm T}({\bf r},{\bf
r'})\cdot {\bf X}^{\rm L}({\bf r'}) & = &  0, \label{propsdeltaeps2}\\
\int\mbox{d}{\bf r'}\; \varepsilon({\bf r'}) {\bf X}^{\rm L}({\bf
r'})\cdot{\bm \delta}_{\varepsilon}^{\rm T}({\bf r'},{\bf r}) & =
& 0,\label{propsdeltaeps3}
\end{eqnarray}
\end{subequations}
for any  transverse function ${\bf X}^{\rm T}$ (zero divergence)
and longitudinal function ${\bf X}^{\rm L}$ (zero curl).

With the use of Eqs.~(\ref{propsdeltaeps1}) and
(\ref{propsdeltaeps3})  the functional derivative
(\ref{funcdifdefgentrans}) is simple in the following two
important cases:
\begin{subequations}
\begin{eqnarray}\label{simplefuncdef}
\frac{\delta}{\delta {\bf Y}_{\varepsilon}^{\rm T}({\bf r})}\int
\mbox{d}{\bf r'}\;{\bf X}^{\rm T}({\bf r'})\cdot {\bf Y}({\bf r'})
& =
& {\bf X}^{\rm T}({\bf r}), \label{firstimportantcase}\\
\frac{\delta}{\delta {\bf Y}_{\varepsilon}^{\rm T}({\bf r})}\int
\mbox{d}{\bf r'}\;\varepsilon({\bf r'}){\bf X}^{\rm L}({\bf
r'})\cdot {\bf Y}({\bf r'}) & = & 0, \label{secondimportantcase}
\end{eqnarray}
\end{subequations}
where ${\bf X}^{\rm T}$ and ${\bf X}^{\rm L}$ are arbitrary
transverse and longitudinal functions, respectively. The second
case (\ref{secondimportantcase}) makes clear that the constrained
functional derivative of inner products of transverse and
longitudinal fields indeed gives zero; the first case
 (\ref{firstimportantcase}) shows that the partial-derivative-of-$\mathcal{F}$ computation
rule still gives the correct answers for inner products of ${\bf
Y}_{\varepsilon}^{\rm T}$ with transverse functions.

How can the constrained functional derivative be calculated in the
more general situation
\begin{equation}\label{funcdifmoregeneral}
\frac{\delta}{\delta {\bf Y}_{\varepsilon}^{\rm T}({\bf r})}\int
\mbox{d}{\bf r'}\;{\bf X}({\bf r'})\cdot {\bf Y}({\bf r'}),
\end{equation}
where ${\bf X}$ is a general vector function? It will now be shown
that any vector field ${\bf X}$ can be decomposed such that the
only two rules of computation needed are the simple cases
(\ref{firstimportantcase}) and (\ref{secondimportantcase}).

Given the vector field ${\bf X}$, construct the scalar field
$\sigma_{\bf X}\equiv -{\bm \nabla}\cdot {\bf X}$. Now find the
potential $\chi$, given the ``charge distribution'' $\sigma_{\bf
X}$ and the dielectric function $\varepsilon_{0}\varepsilon({\bf
r})$, from the following generalized Poisson equation:
\begin{equation}\label{electrostatical}
\varepsilon_{0}{\bm \nabla}\cdot\left[\;\varepsilon({\bf r}){\bm
\nabla}\chi({\bf r})\right] = -\sigma_{\bf X}({\bf r}).
\end{equation}
This is a well-known problem in electrostatics.  There is a unique
solution for $\chi$ of this inhomogeneous problem, given the
charge distribution and the boundary condition that the potential
be zero at infinity. With the potential $\chi$ thus found, define
two vector fields ${\bf X}_{1}$ and ${\bf X}_{2}$ as
\begin{subequations}
\begin{eqnarray}\label{x1x2funcdef}
{\bf X}_{1}  &\equiv &  {\bf X}({\bf r}) -
\varepsilon_{0}\varepsilon({\bf r}){\bm \nabla}\chi({\bf r}),
\label{x1funcdef} \\
{\bf X}_{2} & \equiv & \varepsilon_{0}\varepsilon({\bf r}){\bm
\nabla}\chi({\bf r}), \label{x2funcdef}
\end{eqnarray}
\end{subequations}
so that evidently ${\bf X} = {\bf X}_{1}+{\bf X}_{2}$.  The vector
field ${\bf X}_{1}$ is transverse by construction of the potential
$\chi$; the field $[{\bf X}_{2}/\varepsilon({\bf r})]$ is of
course longitudinal. In summary, the following theorem was proven:
{\em An arbitrary vector field can be uniquely decomposed into a
 part which after division by $\varepsilon({\bf r})$ is longitudinal, and a transverse part.}
 This theorem is useful for evaluating the constrained functional
 derivative, because it leads to
\begin{equation}\label{funcdefdecompo}
\frac{\delta}{\delta {\bf Y}_{\varepsilon}^{\rm T}({\bf r})}\int
\mbox{d}{\bf r'}\;{\bf X}({\bf r'})\cdot {\bf Y}({\bf r'})  =
   {\bf X}({\bf r}) - \varepsilon_{0}\varepsilon({\bf r}){\bm
\nabla}\chi({\bf r}),
\end{equation}
where the unique decomposition of ${\bf X}$ was used and the
simple derivatives Eqs.~(\ref{firstimportantcase}) and
(\ref{secondimportantcase}) were applied to ${\bf X}_{1}$ and
${\bf X}_{2}$, respectively. The problem of computing a functional
derivative with respect to a generalized transverse function has
thus been reduced to a problem in electrostatics. Note that the
constrained functional derivative (\ref{funcdefdecompo}) produces
a  field that is always transverse. That transverse field is
equal to the transverse part of ${\bf X}$ if ${\bf X}$ itself is
transverse or if $\varepsilon({\bf r})\equiv 1$.

The unique decomposition (\ref{x1funcdef}) and (\ref{x2funcdef})
of vector fields is a generalization of the Helmholtz theorem
\cite{Morse53}, but the name ``generalized Helmholtz theorem'' was
already given to a slightly different statement \cite{Dalton97},
namely {\em Every vector field ${\bf Z}$ can be uniquely
decomposed as the sum of a generalized transverse vector field
${\bf Z}_{1}$ and a longitudinal field ${\bf Z}_{2}$}. (The proof
of this theorem in \cite{Dalton97}  begs the question whether the
part of the decomposition that is called longitudinal indeed has
zero curl, but one can show that this is the case.) As a corollary
of the decomposition (\ref{x1funcdef}) and (\ref{x2funcdef}), a
new and short proof can be given of the generalized Helmholtz
theorem. The proof is simple: given ${\bf Z}$, define ${\bf
X}=\varepsilon({\bf r}){\bf Z}$. Then apply the previous unique
decomposition to ${\bf X}$, as in Eqs.~(\ref{x1funcdef}) and
(\ref{x2funcdef}). Define ${\bf Z}_{1}\equiv {\bf
X}_{1}/\varepsilon({\bf r})$ and ${\bf Z}_{2}\equiv {\bf
X}_{2}/\varepsilon({\bf r})$. Then it follows that ${\bf Z}={\bf
Z}_{1}+{\bf Z}_{2}$, where ${\bf Z}_{1}$ is a generalized
transverse and ${\bf Z}_{2}$ is a longitudinal field. This
completes the proof.

\subsection{Functional derivatives of the minimal-coupling
Lagrangian}\label{funcdifminimal} The definition of the
constrained functional derivative and its computation rules can
now be applied to our case of interest, where the functional is
the Lagrangian ${L'}_{\rm min}$, Eq.~(\ref{Lrewritten}), and where
the generalized Coulomb gauge applies to  the vector potential
${\bf A}$.

Before choosing the gauge, the ``ordinary'' functional derivative
(\ref{funcdifdef}) of the Lagrangian of Eq.~(\ref{Lagdensity})
with respect to ${\bf A}$ leads to the fourth Maxwell
equation~(\ref{Maxwellfourth}), as it should. After choosing the
gauge, the interaction term $L_{{\bf A}\Phi}$ [Eq.~(\ref{LAPhi})]
in the Lagrangian becomes identically zero. Its functional
derivative with respect to ${\bf A}$ should also be zero. This is
indeed the case, because the constrained derivative
(\ref{funcdifdefgentrans}) is the correct one to use rather than
the ordinary functional derivative (\ref{funcdifdef}) after
choosing the gauge. Note that the constrained functional
derivative must also be used for free space after choosing the
Coulomb gauge, with ${\bm \delta}_{\varepsilon}^{\rm T}$ equal to
${\bm \delta}^{\rm T}$, Eq.~(\ref{deltatransfree}). For free space
the machinery of functional derivatives usually is not introduced
and the derivative is taken implicitly, for example in
\cite{Cohen87} (p.~289).

The naive calculation of the canonical field in
Eq.~(\ref{canconj}) of section \ref{inhomwithout} can now be
justified:
\begin{equation}\label{funcdifepsakwadraat}
\frac{\delta}{\delta {\bf \dot{A}}_{\varepsilon}^{\rm T}({\bf
r})}\int \mbox{d}{\bf r'}\;\varepsilon({\bf r'}) {\bf
\dot{A}}^{2}({\bf r'})= 2\; \varepsilon({\bf r}){\bf \dot{A}}({\bf
r}).
\end{equation}
One can find this result by realizing that the functional on the
left-hand side is a special case of Eq.~(\ref{firstimportantcase})
with the fields ${\bf X}$ and  ${\bf Y}$ equal to
$\varepsilon({\bf r})\dot{\bf A}$ and $\dot{\bf A}$, respectively.

 In the special case that the
vector field ${\bf X}$ is the current density ${\bf J}_{\rm g}$
produced by the guest atoms in the dielectric, current
conservation implies that the scalar field $\sigma_{\bf J_{\rm
g}}$ as constructed in section~(\ref{simplerules}) equals the
time-derivative of the physical charge density $\sigma_{\rm g}$.
By the uniqueness of the solution of the generalized Poisson
equation, the potential $\chi$ must then be identified with the
time-derivative of the physical scalar potential $\Phi$.
Therefore, the constrained functional derivative of $\int
\mbox{d}{\bf r}\;{\bf J}_{\rm g}\cdot{\bf A}$ can now be computed
as
\begin{equation}\label{funcdifJA}
\frac{\delta}{\delta {\bf A}_{\varepsilon}^{\rm T}({\bf r})}\int
\mbox{d}{\bf r'}\;{\bf J}_{\rm g}({\bf r'})\cdot {\bf A}({\bf
r'})= {\bf J}_{\rm g}({\bf r}) - \varepsilon_{0}\varepsilon({\bf
r}){\bm \nabla}\dot{\Phi}({\bf r}).
\end{equation}
With this result, the Euler-Lagrange equation for the vector
potential from the Lagrangian (\ref{Lrewritten}) precisely becomes
the equation~(\ref{waveforA}) for the vector potential that we
were looking for. Only by the careful computation of functional
derivatives as presented in Secs.~\ref{funcdefdef} and
\ref{simplerules} can one prove that the fourth Maxwell
equation~(\ref{Maxwellfourth}) holds also after choosing the
generalized Coulomb gauge. Interestingly, before choosing the
gauge, the source term $-\varepsilon_{0}\varepsilon({\bf r}){\bm
\nabla} \dot{\Phi}$ in (\ref{waveforA}) came from the $L_{{\bf
A}\Phi}$ interaction term (\ref{LAPhi}) in the Lagrangian. After
choosing the gauge, however, the source term is produced by the
constrained functional derivative of the minimal-coupling
interaction term $\int\mbox{d}{\bf r}\; {\bf J}_{\rm g}\cdot{\bf
A}$.

The left-hand side of the wave equation (\ref{waveforA}) is
certainly transverse in the generalized Coulomb gauge. The
right-hand side is also transverse. A mathematical reason is that
the wave equation is found by functional differentiation with
respect to generalized transverse functions. In
section~\ref{simplerules} it was shown that these derivatives are
always transverse. Physically, the source term must be transverse
because of current conservation. [Use Eq.~(\ref{generalpoisson})].

\section{The quantum multipolar interaction
Hamiltonian}\label{secquantmultiham}

In the previous sections it was shown that the minimal-coupling
Lagrangian produces the Maxwell-Lorentz equations for the
electromagnetic fields and the guest charges, before and also
after choosing the generalized Coulomb gauge. Now the goal is to
transform the gauge-dependent minimal-coupling Lagrangian
(\ref{Lrewritten}) in order to obtain a Lagrangian with multipole
interaction between the electromagnetic field and the guest atoms.
The multipolar Lagrangian must lead to the same equations for the
fields and charges, of course. Candidate transformations are
transformations where a total time derivative of a function of the
canonical variables is added to the Lagrangian. Such
transformations leave the action unchanged \cite{Cohen87}. A
particular transformation of this sort will be used shortly, but
first some new fields must be introduced.

\subsection{Polarization, magnetization and displacement fields}
In the following, it is useful to describe the guest atoms in
terms of a polarization density ${\bf P}_{\rm g}$ and a
magnetization density ${\bf M}_{\rm g}$, rather than in terms of
the charge and current densities. The former and latter pairs are
related through \cite{Jackson75}:
\begin{equation}\label{Pgdef}
\sigma_{\rm g} = -{\bm \nabla}\cdot {\bf P}_{\rm g}, \qquad{\bf
J}_{\rm g}  =  {\bf \dot{P}}_{\rm g} + {\bm \nabla}\times {\bf
M}_{\rm g}.
\end{equation}
In terms of the new variables, the equation of continuity is
automatically satisfied. The polarization and magnetization fields
have the  integral representations \cite{Jackson75}
\begin{widetext}
\begin{subequations}
\begin{eqnarray}\label{pandm}
{\bf P}_{\rm g}({\bf r},t) & = & -e\sum_{mj}
\int_{0}^{1}\mbox{d}u\;({\bf r}_{mj}-{\bf R}_{m})\;
\delta({\bf r}-{\bf R}_{m}-u({\bf r}_{mj}-{\bf R}_{m})\;), \label{pandm1}\\
{\bf M}_{\rm g}({\bf r},t) & = & -e\sum_{mj}
\int_{0}^{1}\mbox{d}u\;u\;({\bf r}_{mj}-{\bf R}_{m})\times {\bf
\dot{r}}_{mj}\; \delta({\bf r}-{\bf R}_{m}-u({\bf r}_{mj}-{\bf
R}_{m})\;). \label{pandm2}
\end{eqnarray}
\end{subequations}
\end{widetext}
These polarization and magnetization  fields are simply the sums
of the  fields ${\bf P}_{{\rm g}m}$ and ${\bf M}_{{\rm g}m}$
produced by the individual guest atoms. Finite-order multipole
expansions of the polarization and magnetization fields can be
found by truncating the Taylor expansion in $u$ of the integrands
on the right-hand sides of the above equations. Such
approximations will be made in section~\ref{Hquantdipoleapp}.

The displacement field ${\bf D}$ and the magnetic field vector
${\bf H}$ are given by the constitutive relations
\begin{equation}\label{displacementwithg}
{\bf D}  \equiv \varepsilon_{0}\varepsilon({\bf r}){\bf E}+  {\bf
P}_{\rm g}, \qquad{\bf H}  \equiv   \mu_{0}^{-1}{\bf B}-{\bf
M}_{\rm g}.
\end{equation}
The electric field ${\bf E}$ and magnetic field ${\bf B}$ are
again defined by Eq.~(\ref{EBintermsofAPhi}) in terms of a scalar
potential $\Phi$ and a vector potential ${\bf A}$. It was assumed
as before that the dielectric is nonmagnetic so that the magnetic
permeability equals the value $\mu_{0}$ of free space. Note that
the dielectric function $\varepsilon({\bf r})$ is a property of
the dielectric alone, independent of the guest atoms. On the other
hand, the displacement field (\ref{displacementwithg}) does depend
on the guest atoms because it includes the polarization field
produced by them. The displacement field defined here is therefore
different (although the same symbol is used)  from the
displacement field that was defined in
section~\ref{chapmuemudclasham} where no guest atoms were present.
In that case,  the transversality of ${\bf D}$ was evident. The
displacement field is also transverse when guest atoms are
present, according to Eqs.~(\ref{Maxwellthird}) and (\ref{Pgdef})
in combination with (\ref{displacementwithg}).

\subsection{Classical multipolar Lagrangian and Hamiltonian}\label{transtomultipole}
The minimal-coupling Lagrangian (\ref{Lrewritten}) will now be
transformed to a multipolar Lagrangian by adding to it the total
time derivative
\begin{equation}\label{transPZW}
-\frac{\mbox{d}}{\mbox{d}t}\int\mbox{d}{\bf r}\;{\bf P}_{\rm
g}({\bf r},t)\cdot {\bf A}({\bf r},t).
\end{equation}
This is the Power-Zienau-Woolley (PZW) transformation, and its
effect is well-known for free space \cite{Cohen87}. The reason to
choose this transformation will be given {\em a posteriori}, when
discussing the multipolar Hamiltonian. The difference with the
free-space PZW transformation is now that the vector potential
satisfies the generalized rather than the usual Coulomb gauge. The
transformation was already applied to inhomogeneous dielectrics
before, in \cite{Dalton96}. There it was stated that the
polarization density ${\bf P}_{\rm g}$ in the Lagrangian density
 should be replaced by a ``reduced polarization
density'' at this point, in order to stick to the generalized
Coulomb gauge for the vector potential. However, as will be clear
shortly, such replacements are not necessary if functional
derivatives with respect to the generalized transverse vector
potential are identified as constrained functional
differentiations, and if computation rules as presented in section
\ref{appfuncdef} are used accordingly.

After adding the  term (\ref{transPZW}), the new Lagrangian
$L_{\rm multi}$ can be rewritten as
\begin{equation}\label{Lprime}
L_{\rm multi} = \sum_{mj}\frac{1}{2}m_{\rm e}{\bf
\dot{r}}_{mj}^{2} -V_{C} + \int\mbox{d}{\bf r}\;\mathcal{L}_{\rm
multi}.
\end{equation}
The new Lagrangian density $\mathcal{L}_{\rm multi}$ has the form
\begin{equation}\label{Ldensityprime}
\mathcal{L}_{\rm multi}  =
\frac{1}{2}\varepsilon_{0}\varepsilon({\bf r}){\bf \dot{A}}^{2} -
\frac{1}{2\mu_{0}}({\bm \nabla}\times {\bf A})^{2} +{\bf M}_{\rm
g}\cdot {\bm \nabla}\times{\bf A} - {\bf P}_{\rm g}\cdot{\bf
\dot{A}},
\end{equation}
where the definition of the magnetization density (\ref{pandm2})
was used as well as Gauss's theorem. The derivation is identical
to the free-space case.

In order to find a Hamiltonian, first the canonically conjugate
variables must be determined. By  reasoning as in section
\ref{funcdifminimal}, the constrained functional differentiation
of the Lagrangian $L_{\rm multi}$ with respect to ${\bf \dot{A}}$
produces the following field that is canonically conjugate to the
vector potential:
\begin{equation}\label{pimultipolar}
{\bm \Pi}  \equiv  \frac{\delta L_{\rm multi}}{\delta\dot{\bf
A}_{\varepsilon}^{\rm T}} = \varepsilon_{0}\varepsilon({\bf
r}){\bf \dot{A}} - [{\bf P}_{\rm g} -
\varepsilon_{0}\varepsilon({\bf r}){\bm \nabla}\Phi] =  -{\bf D}.
\end{equation}
Here,  the definition of the electric field
(\ref{EBintermsofAPhi}) and the displacement field
(\ref{displacementwithg}) were used. As in the case without guest
atoms, the field canonically conjugate to the vector potential
equals minus the displacement field. The difference is that now
the displacement field also contains the polarization field
produced by the guest atoms. The canonically conjugate field would
have been different if the minimal-coupling Lagrangian had been
used.

The result Eq.~(\ref{pimultipolar}) that the canonically conjugate
field of the vector potential is the full displacement field of
the medium including guest atoms, is an important generalization
of the free-space result \cite{Cohen87}. In our formalism, it
could be found rather easily, by realizing that functional
derivatives must be redefined  after choosing a gauge. In
\cite{Dalton96}, a canonically conjugate field was identified that
was stated to be different from the displacement field; in
\cite{Dalton97} the matter was reconsidered and the displacement
field was found as the canonically conjugate field after all, but
only because the polarization field ${\bf P}_{\rm g}$ in the PZW
transformation (\ref{transPZW}) was replaced by a `reduced
polarization field' for reasons that remain somewhat unclear. The
effect of the replacement seems to be that functional derivatives
with respect to the vector potential can be calculated as partial
derivatives, a computation rule that in general is valid only
before choosing the gauge. In contrast, our PZW
transformation~(\ref{transPZW}) features the usual polarization
field of the guest atoms, whether we choose to do the
transformation before or after fixing the gauge. We think that our
approach is more transparent and wider applicable.

The canonical momenta ${\bf p}_{mj}$ corresponding to the
coordinate variables ${\bf q}_{mj}$ of the guest charges are
\begin{equation}\label{pmjwithguests}
{\bf p}_{mj} = m_{\rm e}{\bf \dot{r}}_{mj} - {\bf F}_{mj},
\end{equation}
where the field ${\bf F}_{mj}$ stems from the magnetization
density (\ref{pandm2}) and is defined as
\begin{equation}\label{Fmjdef}
{\bf F}_{mj} \equiv e\int_{0}^{1}\mbox{d}u\;u\;{\bf B}[{\bf
R}_{m}-u\;({\bf r}_{mj}-{\bf R}_{m})]\times({\bf r}_{mj}-{\bf
R}_{m}).
\end{equation}
Note that unlike ${\bf A}$ and ${\bf q}_{mj}$, their canonically
conjugate variables ${\bm \Pi}$ and ${\bf p}_{mj}$ are not fully
electromagnetic or fully atomic in nature, respectively.

All canonical momenta have now been determined, so that  the
multipolar Hamiltonian $H_{\rm multi}$ can be given in terms of
the canonical variables $({\bf r}_{mj},{\bf p}_{mj})$ and
canonical fields $({\bf A},{\bm \Pi})$:
\begin{eqnarray}\label{multipolarhamiltonian}
H_{\rm multi} & = & \sum_{mj} {\bf p}_{mj}\cdot{\bf \dot{r}}_{mj}
+ \int\mbox{d}{\bf r}\;{\bm \Pi}\cdot{\bf \dot{A}} - L_{\rm multi}
 \\ & = &
 H_{\rm rad} + H_{\rm at}+ \sum_{m}\left[ V_{\rm P}^{(m)} + V_{\rm M}^{(m)}\right].
\label{multipolarhamiltonianB}
\end{eqnarray}
The total Hamiltonian consist of a radiative and an atomic part,
plus electric and magnetic interactions between field and matter.
The radiative part  of the Hamiltonian is
\begin{equation}\label{Hraddef}
H_{\rm rad}= \int\mbox{d}{\bf r}\;\left[\frac{{\bm
\Pi}^{2}}{2\varepsilon_{0}\varepsilon({\bf r})} ) +\frac{{\bf
B}^{2}}{2\mu_{0}} \right],
\end{equation}
consisting of an electric and magnetic field-energy term,
respectively. The form of the radiative Hamiltonian has not
changed after adding the guest atoms,  but there is a slight shift
in its interpretation, since the polarization of the atoms is
included in the conjugate field. The atomic polarization field of
atom $m$ interacts with the electromagnetic field as described by
\begin{equation}\label{VPDEF}
V_{\rm P}^{(m)} = \int\mbox{d}{\bf r}\; \frac{{\bf P}_{{\rm
g}m}\cdot {\bm \Pi}}
 {\varepsilon_{0}\varepsilon({\bf r})}.
\end{equation}
In most cases this is the dominant interaction between field and
matter. The usually weaker magnetic interaction consists of two
terms:
\begin{equation}\label{VMdef}
V_{\rm M}^{(m)} =
 - {\bf M}_{{\rm g}m}^{'}\cdot {\bf B}
   +  \sum_{mj}\frac{ F_{mj}^{2}}{2 m_{\rm e}}.
   \end{equation}
The first term is linear in the magnetic field and represents the
paramagnetic energy. Instead of the magnetization ${\bf M}_{{\rm
g}m}$, a reduced magnetization ${\bf M}_{{\rm g}m}^{'}$ has been
used in this  first term. The reduced magnetization is defined as
the magnetization [see equation (\ref{pandm2})] with the ${\bf
\dot{r}}_{mj}$ replaced by ${\bf p}_{mj}/m_{\rm e}$
\cite{Cohen87,Dalton96}. The difference has been corrected for by
a sign change of the second term, which is quadratic in the
magnetic field. This is the diamagnetic energy of the guest atoms
in the nonmagnetic dielectric. It  can be safely ignored from now
on since it is much smaller than the other two interactions (see
\cite{Loudon83}, Sec.~8.6).

Only the atomic part of the Hamiltonian
(\ref{multipolarhamiltonianB}) must still be discussed. It has the
form
\begin{equation}\label{Hatomicdef}
 \sum_{mj}\frac{p_{mj}^{2}}{2 m_{\rm e}}
 + \int\mbox{d}{\bf r}\;\frac{{\bf P}_{\rm
g}^{2}}{2\varepsilon_{0}\varepsilon({\bf r})}
\end{equation}
The first term in the atomic Hamiltonian (\ref{Hatomicdef})
represents the kinetic energy of the guest charges; the second
term is the potential energy of the guest atoms, expressed as a
polarization energy. The Coulomb term $V_{C}$ is absent in the
Hamiltonian, because it cancels against the other term quadratic
in ${\bm \nabla}\Phi$ that one gets when solving
Eq.~(\ref{pimultipolar}) for ${\bf \dot{A}}$ and substituting the
result in the Hamiltonian (\ref{multipolarhamiltonianB}).

It is natural to split the polarization energy in
Eq.~(\ref{multipolarhamiltonianB}) into an intra-atomic and an
interatomic polarization energy, respectively
\cite{Cohen87,Dalton97}:
\begin{equation}\label{interandintra}
\int\mbox{d}{\bf r}\;\frac{{\bf P}_{\rm
g}^{2}}{2\varepsilon_{0}\varepsilon({\bf r})} = \sum_{m}
\int\mbox{d}{\bf r}\;\frac{ P_{{\rm
g}m}^{2}}{2\varepsilon_{0}\varepsilon({\bf r})}+ \sum_{m \neq n}
\int\mbox{d}{\bf r}\;\frac{{\bf P}_{{\rm g}m}\cdot{\bf P}_{{\rm
g}n}}{\varepsilon_{0}\varepsilon({\bf r})}.
\end{equation}
The intra-atomic polarization energy is the potential energy that
keeps an atom together; the interatomic polarization energy is the
only interaction term between neutral atoms in the multipolar
Hamiltonian (\ref{multipolarhamiltonianB}). The Hamiltonian is
still classical, so that the polarization field ${\bf P}_{{\rm
g}m}({\bf r})$ (\ref{pandm1})  is identically zero outside the
smallest sphere surrounding all charges that make up the (neutral)
atom $m$. For that reason,  the classical interatomic polarization
energy (also known as contact energy) is identically zero unless
bounding spheres of distinct guest atoms overlap. In the quantum
mechanical description that will be given shortly, the expectation
value of the polarization energy will not be identically zero. The
atomic wave functions of distinct guest atoms have a non-vanishing
overlap. However, the overlap  falls off exponentially with
interatomic distance and is negligible unless the  distance is of
the order of the size of the atoms.

In the rest of this paper, the guest atoms are assumed more than a
few nanometers apart and their contact energies are neglected.
Then the atomic Hamiltonian for all guest atoms is simply the sum
of single-atom Hamiltonians
\begin{equation}\label{Hatmdef} H_{\rm at}^{(m)} =
\sum_{j}\frac{p_{mj}^{2}}{2 m_{\rm e}}
 + \int\mbox{d}{\bf r}\;\frac{{\bf P}_{{\rm
g}m}^{2}}{2\varepsilon_{0}\varepsilon({\bf r})}. \end{equation}
The total Hamiltonian Eq.~(\ref{multipolarhamiltonianB}) becomes
\begin{equation}\label{quantummultiham}
H_{\rm multi} = H_{\rm rad}+ \sum_{m} \left[\; H^{(m)}_{\rm at} +
V_{\rm P}^{(m)} + V_{\rm M}^{(m)} \;\right].
\end{equation}
Just like in the free-space case \cite{Cohen87}, in the multipole
Hamiltonian (\ref{quantummultiham}) for inhomogeneous dielectrics
there is no instantaneous interaction term left between
well-separated neutral guest atoms. This means that in the
multipolar representation atoms only notice each other because
they interact with the same  (retarded) electromagnetic fields
${\bf D}/[\varepsilon_{0}\varepsilon({\bf r})]$ and ${\bf B}$. Of
course, the multipolar and the minimal-coupling representation
should give identical physical predictions; in
\cite{Craig84,Cohen87} the equivalence is proved for several
observables in free space. The absence of direct interatomic
interactions often makes  calculations simpler in the multipolar
representation. This  justifies the choice of the PZW
transformation (\ref{transPZW}) out of  many candidate
transformations.

\subsection{Quantum multipolar Hamiltonian}
\label{modesandquanthamg}

The goal is now to rewrite  the Hamiltonian
(\ref{quantummultiham}) into a second-quantization form, where
macroscopic quantization has been applied to the electromagnetic
field and microscopic quantization to the guest atoms.

First start with the atomic Hamiltonian, Eq.~(\ref{Hatmdef}).
Following (standard) quantum mechanics, the electron coordinates
${\bf r}_{mj}(t)$ and their canonical momenta ${\bf p}_{mj}(t)$
(\ref{pmjwithguests}), as well as the polarization  field ${\bf
P}_{\rm g}$, become  operators  that work on the atomic wave
functions. The single-atom wave functions can be expanded in terms
of eigenfunctions (labelled $k$) of the single-atom Hamiltonian:
\begin{equation}\label{atomicwavefunction}
\Psi_{m}({\bf r}_{1}, {\bf r}_{2},\ldots,{\bf r}_{Z_{m}};t) =
\sum_{k} c_{mk}(t)\psi_{mk}({\bf r}_{1}, {\bf r}_{2},\ldots,{\bf
r}_{Z_{m}}). \end{equation} Second-quantization notation can now
be introduced by promoting the probability amplitudes $c_{mk}(t)$
and $c_{mk}^{*}(t)$  to become annihilation and creation operators
with standard anti-commutation relations. These operators  become
the atomic canonical variables in the second-quantization picture.
(One could even go back and start with a Lagrangian that
identifies $\Psi_{m}$ and $\Psi_{m}^{*}$ as canonical conjugates
\cite{Power83}.) The atomic operators can be written as sums over
matrix elements. For example, the atomic Hamiltonian of atom $m$
in standard second-quantization notation is $H_{\rm at}^{(m)} =
\sum_{k} E_{k}^{(m)}\;c^{\dag}_{mk} c_{mk}$.

The vector potential was again chosen generalized transverse and
its canonically conjugate field (\ref{pimultipolar}) turned  out
to be transverse again, so that these field operators can be
expanded in terms of generalized transverse modes as in
 Eqs.~(\ref{Anormalmode3}) and (\ref{pinormalmode3}). The
 creation and annihilation operators are written as
$a_{\lambda}^{\dag}$ and $a_{\lambda}$ now that atoms are present.
The radiative part (\ref{Hraddef}) of the Hamiltonian becomes
$H_{\rm rad} = \sum_{\lambda} \hbar
\omega_{\lambda}(a_{\lambda}^{\dag}a_{\lambda}+\frac{1}{2})$.

The interaction terms $V_{\rm P}^{(m)}$  and $V_{\rm M}^{(m)}$ in
second-quantization notation become
\begin{widetext}
\begin{subequations}
\begin{eqnarray}
V_{\rm P}^{(m)}  & = &    -\int\mbox{d}{\bf r}\;\frac{{\bf
P}_{{\rm g}m}({\bf r}) \cdot{\bf D}({\bf
r})}{\varepsilon_{0}\varepsilon({\bf r})}   =
-i\sum_{\lambda}\sum_{k,k'}
\sqrt{\frac{\hbar\omega_{\lambda}}{2\varepsilon_{0}}} \left[
a_{\lambda}c^{\dag}_{mk}c_{mk'}\int\mbox{d}{\bf r}\;{\bf P}_{{\rm
g}m,kk'}({\bf r})\cdot {\bf f}_{\lambda}({\bf r}) -
\mbox{H.c.}\;\right], \label{VPm} \\
 V_{\rm M}^{(m)} & = &  -\int\mbox{d}{\bf r}\;{\bf
M}_{{\rm g}m}^{'}({\bf r}) \cdot{\bf B}({\bf r})   =    -
\sum_{\lambda}\sum_{k,k'}
\sqrt{\frac{\hbar}{2\varepsilon_{0}\omega_{\lambda}}} \biggl\{
a_{\lambda}c^{\dag}_{mk}c_{mk'}\int\mbox{d}{\bf r}\;{\bf M}_{{\rm
g}m,kk'}^{'}({\bf r})\cdot \left[{\bm \nabla}\times{\bf
f}_{\lambda}({\bf r})\right] + \mbox{H.c.} \biggl\}. \label{VMm}
\end{eqnarray}
\end{subequations}
\end{widetext}
 The quantity ${\bf P}_{{\rm
g}m,kk'}({\bf r})$ in Eq.~(\ref{VPm}) is the matrix element of the
polarization field ${\bf P}_{{\rm g}m}({\bf r})$ with respect to
states $\psi_{mk}$ and $\psi_{mk'}$ of atom $m$.  The polarization
field couples to the field $-{\bf
D}/[\varepsilon_{0}\varepsilon({\bf r})]$, which is unequal to
$-{\bf E}/\varepsilon_{0}$. The interpretation of the interaction
is subtle, since in the definition (\ref{displacementwithg}) of
the displacement field the polarization  of the guest atoms is
included. The interaction $V_{\rm P}^{(m)}$ therefore includes a
self-interaction of the polarization field. However, in the
expansion in (\ref{VPm}) of the interaction in terms of the
optical modes, the dielectric function $\varepsilon({\bf r})$
drops out and the coupling becomes rather simple. There are no
analogous self-interactions in the magnetic interaction $V_{\rm
M}^{(m)}$. The magnetic field and the reduced magnetization field
are canonically independent and ${\bf M}_{\rm g}^{'}$ is not
included in the definition of ${\bf B}$. All four terms in the
quantum multipolar Hamiltonian (\ref{quantummultiham}) have now
been given in second-quantization notation.

\subsection{Dipole approximation}\label{Hquantdipoleapp}
An atom is much smaller than an optical wavelength; its spatial
structure can not be probed with light. One can make the
well-known assumption that the polarization and magnetization
fields associated with the atom are concentrated in its center of
mass ${\bf R}_{m}$ (the nucleus, say). Mathematically, this means
that the integrands in Eqs.~(\ref{pandm1}) and (\ref{pandm2}) are
approximated by their values in $u=0$. These two values are the
first terms of two infinite Taylor expansions  in terms of the
variable $u$. The dipole approximation is made by keeping only the
first term.

Incidentally, the next terms in the Taylor expansions would
describe  quadrupole interactions, which can be important when the
guest atoms are not real atoms but other (larger) quantum systems
in interaction with the electromagnetic field.  For example,
quantum dots (``artificial atoms'') are much larger than real
atoms and so their dipole moments can be much larger  as well
\cite{Guest02}. Quadrupole moments are more important for quantum
dots than for real atoms, especially when excited in their near
field by a scanning near-field optical microscope \cite{Zurita02}.
In the following, however,  quadrupole and higher-order moments
are neglected.

 In the dipole approximation, the magnetization
(reduced or not) becomes identically zero and the polarization
field becomes
\begin{equation}\label{Pzeroorderdipoleapp}
{\bf P}_{{\rm g}m}({\bf r})  = \delta({\bf r}-{\bf
R}_{m})\sum_{k,k'}c_{mk}^{\dag}{\bm \mu}_{nk'}^{(m)}c_{mk'},
\end{equation}
where the atomic dipole matrix elements ${\bm \mu}_{kk'}^{(m)}$ of
the guest atom $m$ are defined as
\begin{equation}\label{dipolemomentsdef}
{\bm \mu}_{kk'}^{(m)} \equiv  -e\;\langle \psi_{mk}|\sum_{j} ({\bf
r}_{mj}-{\bf R}_{m})|\psi_{mk'}\rangle.
\end{equation}
With equation (\ref{VPm}), it follows that in the dipole
approximation the interaction energy of an atom with the
electromagnetic field in an inhomogeneous dielectric equals
\begin{widetext}
\begin{equation}\label{VPmdip} V^{(m)}_{\rm dip}
 =  -\sum_{kk'}c_{mk}^{\dag}c_{mk'}{\bm
\mu}^{(m)}_{kk'}\cdot{\bf D}({\bf
R}_{m})/[\varepsilon_{0}\varepsilon({\bf R}_{m})] =
-i\sum_{\lambda}\sum_{kk'}
\sqrt{\frac{\hbar\omega_{\lambda}}{2\varepsilon_{0}}} \left[
a_{\lambda}c_{mk}^{\dag}c_{mk'}\;{\bm \mu}^{(m)}_{kk'}\cdot {\bf
f}_{\lambda}({\bf R}_{m}) - \mbox{H.c.}\;\right].
\end{equation}
\end{widetext}
This gives the important result that inside an inhomogeneous
dielectric, a dipole couples to the field $-{\bf
D}/[\varepsilon_{0}\varepsilon({\bf r})]$. This generalization of
the free-space dipole-coupling \cite{Power83,Ackerhalt84,Cohen87}
was also found in \cite{Dalton96,Dalton97}. As we shall see in the
following subsection \ref{atomicham}, local-field effects can have
a strong influence on this interaction. Still, local-field effects
 are often neglected in macroscopic quantization
theories \cite{Knoell87,Kweon95,Dalton96,Dalton97}.

\subsection{In need of a local-field model}\label{atomicham}
In section \ref{modesandquanthamg}  it was not stressed that the
atomic  Hamiltonian (\ref{Hatmdef}) in general is different for an
atom in a dielectric and in free space. The potential energy in
(\ref{Hatmdef}), which includes the Coulomb potential,  is reduced
by a factor $\varepsilon({\bf r})$ as compared to free space.
(Such a reduction factor is well known for dielectric-filled
capacitors). As a consequence, energy levels and wave functions
will be different in a dielectric. This point is missed if one
starts with a second-quantized description, for example when
introducing in the medium a ``two-level atom'' with known
transition frequency and dipole moment.

First suppose that the dielectric function is a macros\-copically
averaged quantity that does not change on atomic length scales.
Then $\varepsilon({\bf r})$ must be unchanged by introducing a
guest atom and the Hamiltonian (\ref{Hatmdef}) can be approximated
by
\begin{equation}\label{Hatomicepsconst}
H^{(m)}_{\rm at} \approx \frac{1}{2 m_{\rm e}} \sum_{j} p_{mj}^{2}
+ \frac{1}{2\varepsilon_{0}\varepsilon({\bf
R}_{m})}\int\mbox{d}{\bf r}\; P_{{\rm g}m}^{2}.
\end{equation}
If the guest atom were a hydrogen atom, then  its Bohr radius and
dipole moments would increase
 by a factor $\varepsilon({\bf R}_{m})$ and its energy levels
 would be reduced by the same factor, according
to this Hamiltonian. Any visible line in free space would then be
shifted to the infrared in a dielectric. The consequences of
approximation~(\ref{Hatomicepsconst}) would be that the dielectric
has a huge effect on the atom's electronic properties. Now the
reduction of the Coulomb potential  (as well as its screening as a
function of distance) is a well-studied subject in solid-state
physics \cite{Jones85}. Sometimes one finds the full reduction [as
described by the Hamiltonian~(\ref{Hatomicepsconst})], while in
other cases no reduction is found at all.

In general, outer electronic states of atoms will be more affected
by the dielectric than the core electrons. An important reason for
this is the dispersive interaction of the guest atoms with the
atoms that make up the medium. However, such frequency dispersion
in the medium is neg\-lected in the present formalism. We should
therefore not have the ambition to find an atomic Hamiltonian that
leads to correct inner and outer electronic states, including
medium effects. A modest model is needed that meets the
requirement that energy levels taking part in the optical
transitions under study should come out right. Such a model might
be obtained by assuming that the atom sits inside an atom-sized
cavity with a relative dielectric function $\varepsilon({\bf
R}_{m})$ that is constant inside the cavity; in general
$\varepsilon({\bf R}_{m})$ will be different both from the
macroscopic dielectric function just outside the cavity, and
different from unity (the free-space value). More am\-bitious
descriptions of medium effects on atomic Hamiltonians require at
least that dispersion of the dielectric is taken into account,
perhaps starting from a microscopic model of the dielectric
\cite{Knoester89,Ho93,Juzeliunas96}.

A well-known case where reduction of potential energy is important
occurs when doping solid silicon with phosphorus to make an
$n$-type semiconductor. The high dielectric constant of Si
($\varepsilon = 11.7$) reduces the potential energy between the
outermost electron and the rest of the P-atom, so that the
electron can enter the conduction band relatively easily, leaving
a $P^{+}$-ion \cite{Marder00}. As said before, modifications other
than (\ref{Hatomicepsconst}) of the atomic Hamiltonian are
possible. An important example of the other extreme case, where a
reduction of the Coulomb interaction is absent, will be given
shortly.

We are interested in atomic lifetime changes and line shifts
caused by the medium. In general, the medium induces changes both
in the atomic Hamiltonian (\ref{Hatmdef}) and in the atom-field
interactions, as compared to free space. Effects of the medium
that are caused by changes in the atomic Hamiltonian will be
called electronic effects. Changes in atomic dipole moments are an
example of electronic effects. On the other hand, effects due to
modified interactions $V_{\rm P}^{(m)}$ and $V_{\rm M}^{(m)}$
between field and atom will be called photonic effects. It is the
photonic effects, the changes due to altered properties of the
electromagnetic field, which are of primary interest here and in
photonics at large. However, only if the electronic changes of the
atoms are somehow either absent or accounted for, can one study
the photonic effects.  In this respect it is fortunate that  line
shifts due to changes in the interactions (radiative or Lamb
shifts) in a medium usually are too small to be observable and
electronic line shifts dominate. Line shifts can therefore be used
to estimate medium-induced changes in the atomic Hamiltonian.
Given a line shift, one could assign an effective dielectric
function $\varepsilon({\bf R}_{m})$ for the atomic cavity that
produces the observed transition frequency when inserted in the
atomic Hamiltonian Eq.~(\ref{Hatomicepsconst}).

The distinction between  photonic and electronic effects is also
very important in the interpretation of experiments. For example,
the recently observed fivefold reduction of spontaneous-emission
rates inside photonic crystals \cite{Koenderink02b} is a photonic
effect, since possible changes in dipole moments were divided out
by choosing a reference sample with identical electronic effects
\cite{Bechger02}. Some earlier observations of long lifetimes in
photonic crystals must be attributed to electronic effects
\cite{Li01a}.

When studying  photonic effects of a medium, the guest atoms
ideally are electronically the same as in free space, in
particular with the same eigenfrequencies and transition dipole
moments. In that ideal case, that we refer to as the {\em
empty-cavity model}, we have
\begin{equation}\label{emptycavitydef}
\varepsilon({\bf R}_{m}) =1 \qquad\forall m.
\end{equation}
The atomic Hamiltonian is as in the approximation
Eq.~(\ref{Hatomicepsconst}), now with $\varepsilon({\bf R}_{m})$
equal to 1. In other words, guest atoms can only be ideal if the
dielectric function is locally changed to the free-space value 1.
The atom-as-in-free-space sits inside an empty cavity inside the
dielectric. A reduction of the intra-atomic Coulomb interaction is
completely absent in this empty-cavity model. The formation of
such a cavity is beyond the scope of the present macroscopic
theory. This would require microscopic theories of the dielectric,
involving the Pauli exclusion principle for electrons of both the
dielectric and the guest atoms.

 The empty-cavity model captures the
observed absence of large electronic effects of the dielectric on
atomic properties of interest, but at the same time the model has
consequences for photonic properties: the local changes in
$\varepsilon$ will give local changes in the mode functions ${\bf
f}_{\lambda}$, and therefore in the dipole coupling
(\ref{VPmdip}). Atomic spontaneous-emission rates  will get
local-field corrections. These predictions can be tested
experimentally. An important example is the emission rate of an
atom inside an atomic-sized empty cavity in an otherwise
homogeneous medium. To be precise, $\varepsilon({\bf
R})=\varepsilon$ for ${\bf R}$ not coinciding with any of the
${\bf R}_{m}$. The  emission rate  is $[3\varepsilon/(2\varepsilon
+1)]^{2}\sqrt{\varepsilon}\Gamma_{0}$, where $\Gamma_{0}$ is the
free-space emission rate \cite{Glauber91}. The well-known
in-medium enhancement by a factor $\sqrt{\varepsilon}$ is further
enhanced by the square of a so-called local-field factor. Here,
the term between the square brackets is the empty-cavity
local-field factor.

In a recent study \cite{Schuurmans98} of refractive-index
dependent spontaneous emission rates, atoms were embedded in a
low-index molecular complex so as to electronically separate them
from the medium. For the interpretation of the results, it was
important that atomic spectra and  dipole moments did not change
appreciably while varying the refractive index. The  empty-cavity
local-field factor was indeed observed \cite{Schuurmans98}. This
result is a justification for the macroscopic quantization theory
for nondispersive dielectrics.

For inhomogeneous dielectrics, it is in general not easy to
calculate local-field factors, either in the empty-cavity model
(\ref{emptycavitydef}) or in other models.  The simplest
assumption in the empty-cavity model is that the
position-dependent local-field factors will have values
$3\varepsilon_{\rm b}({\bf R}_{m})/(2\varepsilon_{\rm b}({\bf
R}_{m}) +1)$, where $\varepsilon_{\rm b}({\bf R}_{m})$ is the bulk
dielectric function around atom $m$. The assumption will probably
break down when $\varepsilon_{\rm b}({\bf R})$ varies strongly on
the scale of the wavelength of light.

The atomic Hamiltonian can be changed in many ways and
consequently, empty-cavity factors are not the only local-field
factors that can be obtained from the present macroscopic
quantization formalism.  One could give up the macroscopic
quantization as being too phenomenological and instead describe
the microscopic constituents of the dielectric in the vicinity of
the guest atom. This could lead to other local-field factors,
depending on the question whether the guest atom sits inside a
real cavity inside the dielectric (of which the empty cavity
(\ref{emptycavitydef}) is a special case), or not. For homogeneous
dielectrics, see \cite{Knoester89,DeVries98b,Schuurmans00} and
references therein. However, for inhomogeneous dielectrics it will
be hard to tie  a local microscopic approach to the macroscopic
description of the inhomogeneous medium on a larger scale.

\section{Dipole-coupling controversy}\label{dipolecontroversy}
Many papers appeared in the nineteen-eighties about the
equivalence of the minimal-coupling and the multipolar Hamiltonian
in free space, for example
\cite{Power78,Power83,Babiker83,Ackerhalt84,Power85}. The
Hamiltonians sometimes lead to different results in calculations.
Some authors argued that the minimal-coupling Hamiltonian was to
be preferred, while others proposed to refrain from using
gauge-dependent equations to stop the confusion. In the multipolar
picture, a controversy arose whether a dipole in free space
couples to minus the displacement field
 $-{\bm \mu}\cdot{\bf D}/\varepsilon_{0}$,  or to the
transverse part of the electric field $-{\bm \mu}\cdot{\bf E}^{\rm
T}$. The first answer is correct and the book by Cohen-Tannoudji
{\em et al.} helped to settle the argument \cite{Cohen87}. It may
be useful to give two sources of confusion even for an atom in
free space, and to compare the free-space dipole coupling  with
its  in-medium generalization Eq.~(\ref{VPmdip}).

The main source of confusion is related to approximations. It was
found in section~\ref{Hquantdipoleapp} that a dipole couples to
minus the displacement field
\begin{equation}\label{dipkopnoges}
-{\bf D}({\bf R}_{m})/[\varepsilon_{0}\varepsilon({\bf R}_{m})]  =
- \frac{\left[\varepsilon({\bf R}_{m}){\bf E}({\bf R}_{m})+ {\bf
P}_{\rm g}({\bf R}_{m})\right]}{\varepsilon_{0}\varepsilon({\bf
R}_{m})},
\end{equation}
where ${\bf P}_{\rm g}$ is the polarization field of the guest
atom itself. In free space, or when assuming an empty-cavity
model, a dipole couples to the field $-{\bf D}/\varepsilon_{0}$.
Still, one can find references stating that in free space or in a
dielectric \cite{Glauber91} a dipole couples to minus the electric
field. The origin of this mistaken interpretation can be traced
back to the normal-mode expansion of the  displacement field
operator for the dielectric with guest atoms
\begin{equation}\label{Dindielwithatoms}
{\bf D}({\bf r}) =  i \varepsilon_{0}\varepsilon({\bf r})
\sum_{\lambda}\sqrt{\frac{\hbar\omega_{\lambda}}{2\varepsilon_{0}}}
\left[a_{\lambda}\;{\bf f}_{\lambda}({\bf r}) -
a_{\lambda}^{\dag}\;{\bf f}_{\lambda}^{*}({\bf r})\right].
\end{equation}
The expanded form of the displacement operator is almost the same
as in a dielectric without guest atoms [minus
Eq.~(\ref{pinormalmode3})], but the two differences will now be
discussed. The first difference, which also plays a role for free
space, is that the equations of motion of the creation- and
annihilation operators in (\ref{Dindielwithatoms}) have terms
involving the atomic variables, which the equations of motion of
their counterparts $a_{\lambda}^{(0)}$ and $a_{\lambda}^{(0)\dag}$
for the field without guests do not have. If one approximates the
displacement field (\ref{Dindielwithatoms}) by replacing all the
$a_{\lambda}$ and $a_{\lambda}^{\dag}$ by $a_{\lambda}^{(0)}$ and
$a_{\lambda}^{(0)\dag}$, respectively, then in free space the
displacement field (\ref{Dindielwithatoms}) is equal to the
electric field  in the absence of the guest atoms. In other words,
if the guest atoms are taken into account in Maxwell's equations,
then one finds a dipole coupling to the displacement field,
whereas a coupling to the electric field is found when guest atoms
are left out of Maxwell's equations.  It depends on the observable
under study whether the difference between the two dipole
couplings can be neglected or not.

The second  difference between the displacement field
(\ref{Dindielwithatoms}) (with guests) and its counterpart (no
guests) is that mode functions will be changed locally when guest
atoms are present, as discussed previously. This difference does
not show up in the free-space discussion, of course. In a
dielectric, the above approximation of replacing  the creation-
and annihilation operators does not make the displacement field
equal to the electric field. The replacement would only have this
effect for positions ${\bf r}$ in the medium where
$\varepsilon({\bf r})$ equals 1 and where local-field effects can
be neglected.

Apart from  the main source of confusion, there is another reason
why the interpretation of the dipole coupling can be confusing:
 there are two essentially different procedures to
go from a minimal-coupling Hamiltonian to a dipole Hamiltonian.
The first procedure is to rewrite a minimal-coupling Hamiltonian
as a multipolar or dipole Hamiltonian by canonical transformations
of its variables (new variables, same Hamiltonian and states). The
second procedure is a unitary change of picture (new Hamiltonian,
new states, same expectation values). Unlike canonical
transformations, picture changes have no classical analogues.
Confusion is likely to arise when after a canonical change a
Hamiltonian has exactly the same form as after a picture change.
The differences between the two procedures are excellently
presented in \cite{Ackerhalt84} for an atom in free space.  Either
of the two procedures could be chosen for dielectrics as well; in
this paper, the PZW transformation of the Lagrangian was used
instead.

\section{Inhomogeneous magnetic media}\label{magneticeffects}
The present formalism can be generalized to dielectrics with
inhomogeneous magnetic properties as well, where the magnetic
permeability becomes $\mu_{0}\mu({\bf r})$ rather than $\mu_{0}$.
(So here ``$\mu$'' does not represent the magnitude of a dipole
${\bm \mu}$.) Such generalizations are even more interesting now
that so-called left-handed materials \cite{Veselago68} have become
the subject of intense scientific discussions, after a prediction
that a perfect lens could be made with them \cite{Pendry00}. Both
 $\varepsilon({\bf r})$ and  $\mu({\bf r})$ of left-handed
materials are negative. The index of refraction $n({\bf r})$ is
also negative and this leads to many peculiar properties.
Left-handed materials will  influence spontaneous-emission rates
of nearby guest atoms in different ways than their right-handed
counterparts \cite{Klimov02}.

Generalizations to magnetic media were already considered in
\cite{Dalton96,Dalton97} and it is relatively straightforward to
incorporate position-dependent permeabilities in the formalism of
this paper, as we will see now. When $\mu$ becomes
position-dependent, then the only term that will change in the
classical multipolar Hamiltonian (\ref{quantummultiham}) is the
magnetic field energy $\int\mbox{d}{\bf r}\;{\bf B}^{2}({\bf
r})/[2\mu_{0}\mu({\bf r})]$. The quantum mechanical description
can again be carried out by choosing the generalized Coulomb gauge
for the vector potential. Without guest atoms, the vector
potential satisfies the source-free wave equation
\begin{equation}\label{sourcefreemagnetic} {\bm \nabla}\times
\left[\frac{1}{\mu({\bf r})}{\bm \nabla}\times {\bf A}({\bf
r})\right] + \frac{\varepsilon({\bf r})}{c^{2}}\ddot{\bf A}({\bf
r}) = 0.
\end{equation}
The electromagnetic field  can be expanded in terms of new true
modes ${\bf m}_{\nu}$ different from the  modes ${\bf
f}_{\lambda}$. The new modes are the harmonic solutions of the
wave equation (\ref{sourcefreemagnetic}) and so they are
generalized transverse, just like the modes ${\bf f}_{\lambda}$ of
the nonmagnetic medium. Canonical fields can be found by
calculating constrained functional derivatives for the new
Lagrangian in the same way as presented in section
\ref{funcdefdef}. It is this point that makes the generalization
 to magnetic media relatively simple. In
second-quantization notation, the electromagnetic field energy
becomes $\sum_{\nu}\hbar\omega_{\nu} \left(d_{\nu}^{\dag}d_{\nu}
+\frac{1}{2}\right)$, where $d_{\nu}^{\dag}$ is the creation
operator of a photon in the mode ${\bf m}_{\nu}({\bf r})$. The new
modes can have mode profiles that differ much from any of the
modes ${\bf f}_{\lambda}({\bf r})$, but otherwise the theoretical
description of the medium is not much different. In the electric
and magnetic interactions (\ref{VPm}) and (\ref{VMm}) of the
electromagnetic field with guest atoms,  the modes ${\bf
f}_{\lambda}$ can just be replaced by the ${\bf m}_{\nu}$ and the
operators $c_{\lambda}^{(\dag)}$ by $d_{\nu}^{(\dag)}$ in order to
take both the electric and magnetic properties of the medium into
account. As for nonmagnetic media, in the dipole approximation the
magnetic interaction (\ref{VMm}) is zero;
 the electric dipole interaction (\ref{VPm}) dominates, except for
 optical transitions with zero dipole moments. When dipole moments
 are nonzero, the main effect of the dielectric becoming  magnetic
 comes from the change in the mode functions.

\section{Summary and discussion}\label{summarymuemudchap}

 The aim of the paper was to find a
 Hamiltonian of guest atoms in an inhomogeneous dielectric,
 with a multipolar interaction between the atoms and the electromagnetic field.
The multipolar Hamiltonian is simpler than the minimal-coupling
Hamiltonian, because in the former all interactions between the
atoms are mediated by the retarded electromagnetic field. The main
results are therefore the quantum multipolar interaction
Hamiltonian (\ref{quantummultiham}) together with its dipole
approximation (\ref{VPmdip}). With this Hamiltonian, one can study
how an inhomogeneous dielectric environment can change quantum
optical processes of resonant atoms.

      In order to obtain the central results,  first  the electromagnetic field
      was quantized in the absence of guest atoms. This has certainly been carried
out before \cite{Knoell87,Glauber91,Vogel94,Dalton96}, but after
explaining why real optical mode functions can be used whenever
$\varepsilon({\bf r})$ is real, the quantization becomes simpler
than found in \cite{Glauber91,Vogel94,Dalton96}. In particular,
only when real mode functions are chosen are the associated
generalized positions and momenta Hermitian [see
Eq.~(\ref{Apinormalmode})].

Guest atoms were introduced into the theory  such that Maxwell's
equations hold with the atomic charge and current densities as
source terms. A minimal-coupling Lagrangian that gives rise to
these equations was easily written down. However, after choosing a
generalized Coulomb gauge, it was not directly clear how to obtain
all Maxwell's equations in this particular gauge. In section
\ref{appfuncdef}, it was explained that the usual functional
derivative must be replaced by a ``constrained functional
derivative'' after choosing a gauge. This is for mathematical
reasons rather than a matter of taste or convenience. Simple rules
are given to actually compute these constrained functional
derivatives.  As a result, the gauge-independent Maxwell's
equations were found indeed to hold in the generalized Coulomb
gauge as well [see Eq.~(\ref{funcdifJA})].

There is a second advantage of our careful treatment of functional
differentiation. In the multipolar formalism, the field in the
dielectric canonically conjugate to the vector potential could
relatively easily and unambiguously be identified as minus the
full displacement field [see Eq.~(\ref{pimultipolar})]. This field
contains the polarization fields produced by both the dielectric
and the guest atoms.

Another important result is that the macroscopic description of
the dielectric could only be tied up to the microscopic
description of the atoms by  assuming that  the dielectric
function is locally modified by the presence of the guest atoms.
In particular,  in the empty-cavity model
[Eq.~(\ref{emptycavitydef})], the dielectric function has the
value 1 where wave functions of the guest atoms are nonzero. The
local modification of the dielectric function will also change the
dipole coupling, giving rise to local-field effects in
spontaneous-emission rates. If one would start with a two- or
three-level description for the guest atoms, then one implicitly
already assumes  a local-field model for the dielectric function.
Consistency requires to  also choose
 that local-field model when calculating the mode functions  in the
dipole interaction (\ref{VPmdip}).

 The application of the Power-Zienau-Woolley
transformation to the minimal-coupling Lagrangian was shown to
produce the multipolar Lagrangian,  after the generalized Coulomb
gauge had been chosen. Actually, the gauge was chosen earlier than
strictly necessary: the choice could have been postponed until the
canonical momenta were determined from the multipolar Lagrangian.
The story would have been simpler up to that point. The difficulty
to find all Maxwell's equations would then show up only after
obtaining the multipolar Lagrangian. The reason to first choose
the gauge and then do the PZW transformation, is that it more
clearly shows that the difficulty to find all Maxwell's equations
was a consequence of choosing the gauge, rather than a consequence
of the PZW transformation.

More generally, the presentation given here is one among  many
possibilities. When going from a classical minimal-coupling
Lagrangian to a quantum mechanical multipolar Hamiltonian, one has
to make four steps: one step is to choose a gauge, another step is
to transform the theory to the multipolar formalism. Yet another
step is made when going from a Lagrangian to a Hamiltonian;
quantization and second quantization together are step number
four. These are now given in the order in which they occurred in
this paper, but the steps can be interchanged. Not all of the 24
permutations are convenient, but all routes should lead to
equivalent final results. It was shown in detail in
\cite{Babiker83} that step two and three can be interchanged for
free space: the PZW transformation of the minimal-coupling
Lagrangian is equivalent to a picture change of the
minimal-coupling Hamiltonian. The equivalence will also hold for
inhomogeneous dielectrics.

\section*{Acknowledgements}
We would like to thank Allard Mosk, Rudolf Sprik, and Willem Vos
for stimulating discussions. This work is part of the research
program of the Stichting voor Fundamenteel Onderzoek der Materie,
which is financially supported by the Nederlandse Organisatie voor
Wetenschappelijk Onderzoek.

\end{document}